\documentclass[twocolumn]{aastex7}

\usepackage{multirow}
\usepackage{bold-extra}

\newcommand{\odltecolor}[0]{yellow}
\newcommand{\tdltecolor}[0]{red}
\newcommand{\odnltecolor}[0]{green}
\newcommand{\tdnltecolor}[0]{cyan}

\def\one{{\,\sc i}}
\def\two{{\,\sc ii}}
\def\three{{\,\sc iii}}
\def\four{{\,\sc iv}}
\def\five{{\sc v}}
\def\six{{\sc vi}}

\newcommand{\code}[1]{\texttt{#1}}
\newcommand{\cmfgen}{\code{CMFGEN}}

\begin{document}

\title{Non-LTE Synthetic Observables of a Multidimensional Model of Type Ia Supernovae}

\author[0000-0002-1184-0692]{Samuel J. Boos}
\affiliation{Department of Physics \& Astronomy, University of Alabama, Tuscaloosa, AL, USA}
\email{sjboos@crimson.ua.edu}

\author[0000-0003-0599-8407]{Luc Dessart}
\affiliation{Institut d’Astrophysique de Paris, CNRS-Sorbonne Université, 98 bis boulevard Arago, F-75014 Paris, France}
\email{dessart@iap.fr}

\author[0000-0002-9538-5948]{Dean M. Townsley}
\affiliation{Department of Physics \& Astronomy, University of Alabama, Tuscaloosa, AL, USA}
\email{dmtownsley@ua.edu}

\author[0000-0002-9632-6106]{Ken J. Shen}
\affiliation{Department of Astronomy and Theoretical Astrophysics Center, University of California, Berkeley, CA, USA}
\email{kenshen@astro.berkeley.edu}


\begin{abstract}
Many promising explosion models for the elusive origin of Type Ia supernovae (SNe Ia) ultimately fail to completely reproduce a number of observed properties of these events.
One limiting factor for many of these models is the use of the local thermodynamic equilibrium (LTE) assumption in the calculation of their synthetic observables, which has been shown to prevent the accurate prediction of a number of fundamental features of SNe Ia.
The inclusion of high-accuracy non-LTE physics, however, increases computational cost and complexity such that multidimensional non-LTE calculations are often unfeasible, which can be problematic for models that are inherently multidimensional.
In this work, we conduct radiative transfer calculations using 1D profiles that each correspond with a line of sight from an asymmetric, 2D SN Ia model.
We find, in LTE, that the synthetic observables from these calculations efficiently reproduce those from the 2D calculation when an equivalence of bolometric luminosities between the 1D and 2D treatments is enforced.
This allows for the accurate calculation of synthetic observables in 1D while still preserving multidimensional effects associated with the model.
We leverage this to produce high accuracy observables from 1D non-LTE calculations, showing significantly improved agreement with observation, including a roughly 50\% reduction of $B$-band decline rate into congruence with the observed Phillips relation. 
Additionally, our non-LTE observables show Si \textsc{ii} $\lambda$5972 pEWs that are much more similar to observation, while spanning multiple Branch classes, suggesting that some spectral classifications of SNe Ia may arise from line of sight effects.
\end{abstract}

\keywords{Type Ia supernovae --- Radiative transfer}

\section{Introduction}

Type Ia supernovae (SNe Ia) are integral to a number of aspects in astronomy and cosmology, including the chemical evolution of the Universe, discovery of the accelerating expansion of the Universe, and determination of the local Hubble constant \citep{riess_1998,perlmutter_1999}.
While it is general consensus that these events arise from the thermonuclear explosion of a carbon/oxygen white dwarf in a binary system, the exact progenitor systems that undergo a SN Ia, as well as the ignition mechanism and type of explosion, remain uncertain.
A number of proposed models for SNe Ia have shown promise, however none have yet been able to precisely reproduce the observable characteristics of both individual events and the population of SNe Ia as a whole.
A model, or collection of models, that could accurately predict the observed properties of SNe Ia would improve the utility of SNe Ia across numerous areas in astronomy and cosmology.

A wide variety of progenitor systems and explosion models for SNe Ia have been proposed over the last few decades \citep[see][for a recent detailed review]{liu_2023}.
In summary, SNe Ia originate from exploding white dwarfs that may have a mass that is Chandrasekhar or sub-Chandrasekhar and the explosion is triggered via some interaction with its binary companion.
In the case of a Chandrasekhar-mass progenitor, the common view is that the central densities increase during mass transfer from a non-degenerate companion until the white dwarf undergoes a deflagration and/or a detonation \citep{Whelan_1973,Iben_1984,Nomoto_1984,hoeflich_1995,ropke_2007,seitenzahl_2012,sim_2013}.
Alternatively for sub-Chandrasekhar progenitors, a favored scenario is the double detonation, in which the explosion of the white dwarf is triggered by the detonation of its helium-rich shell, which is initiated during mass transfer from its likely degenerate companion \citep{Nomoto_1982,Woosley_1986,fink_2010,polin_2019,Shen_2021_dd,collins_2022,shen_2024}.
It is plausible that both of these progenitor channels are responsible for observed SNe Ia \citep{maoz_2014}.

Many of these proposed models show potential in their synthetic observables with moderate agreement with some individual observed events.
However, broad reproduction of the properties of SNe Ia as a whole remains elusive.
For example, while the double detonation scenario has proven an ability to generate spectra that emulate individual events across a range of peak magnitudes, the Phillips and color-magnitude relations of observed SNe Ia are not yet well-replicated by multidimensional models \citep{Shen_2021_dd,collins_2022}.

Just as the details and treatment of the physics in explosion models are important, so are those when it comes to the radiative transfer to generate synthetic observables.
One commonly used approximation in radiative transfer calculations for SNe Ia is the assumption of local thermodynamic equilibrium (LTE).
Under LTE, certain properties of the SN Ia ejecta are determined by the local gas temperature, including the excitation and ionization levels.
In reality, however, this prescription of equilibrium is not necessarily enforced, and thus non-LTE calculations must be undertaken in order to approach the most physically-realistic radiative transfer processes and, correspondingly, the most accurate synthetic observables.

This has borne out in some recent works that have shown distinct differences in the synthetic observables between LTE and non-LTE calculations of sub-Chandrasekhar detonation models.
\citet{Shen_2021_nlte} showed a notable difference in the spectra and light curves from 1D sub-Chandrasekhar explosion models between LTE and non-LTE calculations beyond maximum light, especially within $U$- and $B$-band.
\citet{blondin_2022} also showed in a radiative transfer code comparison study that \texttt{CMFGEN} \citep{hillier_2012}, a non-LTE code, produces synthetic observables from a sub-Chandrasekhar SN Ia ejecta profile that have generally faster rises to peak and bluer colors compared to most LTE codes.
Additionally, \citealt{collins_2023} identified a non-LTE-exclusive feature, He \textsc{I} $\lambda10830$, which may be a unique signature of the double detonation scenario.
We refer the reader to \citet{dessart_2014_critical} for a detailed exploration into the critical ingredients for non-LTE calculations of a SN Ia model.

While non-LTE is clearly important for the generation of accurate synthetic observables, it is very computationally expensive such that the vast majority of non-LTE calculations for SNe Ia are only performed on 1D profiles.
However, there are a number of SN Ia models that are inherently multidimensional, including the previously mentioned double detonation scenario which sees a stark diversity of observables, including spectral shape and peak magnitude, across lines of sight of a given explosion.
This challenge to generate high accuracy observables from multidimensional models thus conflicts with our ability to compare to these models to the numerous ways that SN Ia are observationally correlated and categorized.
For example, this issue prevents the accurate prediction of the Phillips relation, the primary trend of SNe Ia, from a multidimensional model.

Given the described disparity in observed properties of SNe Ia and what is obtainable from present day calculations of theoretical models, we seek to develop a method in which non-LTE physics is taken into consideration in radiative transfer calculations of a multidimensional SN Ia model, allowing for the generation of accurate synthetic observables from such a model.
In this work, we describe a new method that strives to produce high-accuracy, non-LTE observables from a multidimensional SN Ia model using relatively modest computational resources. 
In short, this method involves a combination of 1D LTE, 2D LTE, and 1D non-LTE radiative transfer calculations to produce ``pseudo-2D'' non-LTE synthetic observables.
We describe our one examined model and radiative transfer methods in Section \ref{sec:methods}.
In Section \ref{sec:results}, we describe the synthetics observables that arise from our calculations and compare them to observed SNe Ia in Section \ref{sec:compare_to_obs}.
Finally, in Section \ref{sec:conclusions}, we discuss our conclusions from this work.


\section{Methods}
\label{sec:methods}
We describe our choice of SN Ia model for this study in Section \ref{subsec:model_choice}.
We also show how we compute the LTE and non-LTE synthetic observables in Sections \ref{subsec:lte_calcs} and \ref{subsec:nlte_calcs}, respectively.
Finally, we describe our strategy to construct pseudo-2D non-LTE observables in Section \ref{subsec:pseudo2d}.

\subsection{Double Detonation Model}
\label{subsec:model_choice}
For this initial work, we probe a single 2D double detonation model from \citet{boos_2021}.
We choose the 1.0 $M_{\odot}$ total mass WD with a 0.016 $M_{\odot}$ helium shell.
This model has a shell base density of $2\times10^{5}$ g cm$^{-3}$, which is the lowest shell density examined in \citealt{boos_2021}.
While the helium shell in the double detonation has historically been thought to necessitate a build up in mass from accretion prior to ignition, \citet{shen_2024} recently showed that the natal shell with a realistic composition profile of a degenerate progenitor up to $\sim 1.0$ M$_{\odot}$ is capable of hosting a detonation, thus we choose the double detonation model with the lowest shell mass available.

\begin{figure*}
    \centering    
    \includegraphics[width=1.0\textwidth]{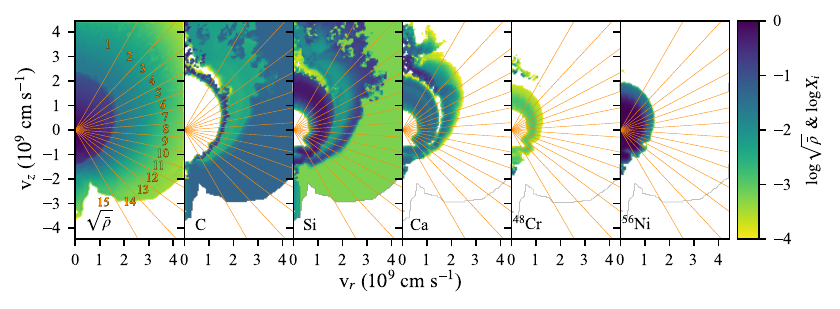}
    \caption{2D density and composition ejecta maps at approximately 100 s post-explosion for the thin shell, 1.0 $M_{\odot}$ total mass WD model from \citet{boos_2021}.
    The orange lines indicate the boundaries of the 15 wedges that are used to construct the 1D profiles for our 1D calculations, which are numbered in orange in the first frame.
    }
    \label{fig:boos21_ejecta}
\end{figure*}

The final ejecta structure in velocity space is displayed in Figure \ref{fig:boos21_ejecta}, showing density and a selection of notable elements and isotopes.
The core ashes are bounded by the innermost layer of carbon, which originates from the outermost layer of the carbon/oxygen core which is not completely burned.
The structure of the core ejecta is what can be expected from most sub-Chandrasekhar detonation models, where the innermost region is comprised of $^{56}$Ni and other iron group elements.
Outward of this region, the ejecta transitions to intermediate mass elements like silicon.
Beyond the core ejecta are the products of the helium detonation, which are dependent on its initial density, with more massive shells producing non-trivial amounts of iron group elements.
The model selected for this work had a fairly thin shell with a peak density of 2$\times10^{5}$ g cm$^{-3}$ at the core-shell interface and total mass of 0.016 M$_{\odot}$.
Thus, the shell ashes are predominately comprised of intermediate mass elements around the core-shell boundary and unburned helium at higher velocities.
This ejecta is a relatively standard expectation for the thin-shelled double detonation scenario, including the stratified core and shell regions as well as their asymmetry, where the velocities are higher in the positive z direction, towards the original helium shell ignition point \citep{fink_2010,tanikawa_2018,gronow_2021,ferrand_2022}.

Drawn over the ejecta in orange in Figure \ref{fig:boos21_ejecta} are the boundaries of the ejecta between which we average to form our 1D profiles, or ``wedges''.
As the 2D ejecta is from a cylindrically symmetric simulation, the wedges are equally spaced in $\cos\theta$ so that each represent a uniform solid angle of the ejecta.
The angles spanned by these wedges in the 2D ejecta correspond to the line of sight bins used in our 2D radiative transfer equations, which therefore each represent an equal probability of being observed.

\subsection{LTE Calculations}
\label{subsec:lte_calcs}
The time dependent 2D LTE calculation for our model of focus from \citet{boos_2021} appears in \citet{Shen_2021_dd} using Sedona \citep{kasen_2006}.
In addition to that 2D calculation, 1D LTE calculations are carried out here using Sedona for each of the 15 wedge profiles constructed from the 2D ejecta (areas between adjacent pairs of orange lines in Figure \ref{fig:boos21_ejecta}).
For each wedge, we generate spectra every 1.5 days between three days before and 15 days after the time of maximum $B$-band magnitude from the corresponding line of sight in the 2D calculation.

A typical 1D radiative transfer calculation using these wedge profiles will generally misestimate the total outgoing luminosity compared to the 2D result at the corresponding line of sight.
This is due to the fact that contributions from adjacent wedges affect the outgoing luminosity for a given line of sight in the 2D calculation.
For example, the northernmost wedge has the lowest amount of $^{56}$Ni and will significantly underestimate the outgoing luminosity in the 1D steady-state treatment due to higher amounts of $^{56}$Ni in nearby wedges that increase the luminosity at the corresponding line of sight in the 2D calculation.
Thus, we elect to conduct these 1D LTE radiative transfer calculations in steady-state, allowing us to modulate the outgoing luminosity at each time step.
Aside from energy contributions from adjacent wedges, a typical steady-state calculation in the photospheric phase will also generally underestimate the total outgoing luminosity due to a lack of consideration of the deposited energy from radioactive decays prior to the calculation time, which is otherwise accounted for in a time-dependent calculation.

To unify the 1D LTE luminosities with corresponding 2D LTE values, we increase the power of our 1D LTE calculations by scaling the average radioactive decay energy in the $\rm{^{56}Ni - ^{56}Co - ^{56}Fe }$ decay chain.
We modulate this scaling factor for each wedge and timestep using a Newton-Raphson iteration, until the outgoing bolometric luminosity of each 1D steady-state calculation is within 0.5\% of that from the corresponding line of sight in the 2D calculation at the same time.
The amount of additional power needed to match the 2D calculation is less at later times due to the larger contribution of the instantaneous radioactive decay luminosity to the outgoing luminosity and decreasing overall opacity of the ejecta.
We show an alternative, but less robust, method to unify the 1D steady-state and 2D time-dependent luminosities in Appendix \ref{sec:alt_power}.
Our imposed constraint on the bolometric luminosity is supported by previous works that show fairly consistent bolometric light curves between LTE and non-LTE calculations of SNe Ia models within a month of the explosion \citep{dessart_2014_critical,Shen_2021_nlte,blondin_2022,sharon_2024}, allowing for bolometric unification between our 1D and 2D calculations.
To avoid confusion with our truly 2D LTE results, we refer to these calculations and resulting synthetic observables as ``1D'' LTE, even though they are enhanced in such a way to ideally replicate a 2D result.


\subsection{Non-LTE Calculations}
\label{subsec:nlte_calcs}

In this section we describe the procedure used for evolving the ejecta structures along each wedge of the 2D explosion model with the radiative transfer code \cmfgen\  \citep{hillier_miller_98,dessart_hillier_05,hillier_2012}. 
We used the standard time-dependent non-LTE solver in \cmfgen\ with an approach and setup analogous to those used in previous Type Ia SN studies (see, e.g., \citealt{blondin_2017}).
A few days before $B$-band maximum, the same wedge structure that is used in Sedona is imported into \cmfgen\ and remapped onto an optical-depth scale counting about 90 depth points and covering from 250 until about 45,000 km s$^{-1}$.
For the composition, we account for the presence of He, C, N, O, Ne, Na, Mg, Al, Si, S, Ar, Ca, Sc, Ti, V, Cr, Mn, Fe, Co, and Ni.
We consider three two-step decay chains with parent isotopes $^{56}$Ni, $^{48}$Cr, and $^{52}$Fe.
Non-thermal rates arising from Compton-scattered, high-energy electrons are computed through a solution to the Spencer-Fano equation, as discussed in \citet{li_etal_12_nonlte} and \citet{d12_snibc}. 

We employ an extended model atom with the following ions, super-levels and full levels (see \citealt{hillier_miller_98} for details about the treatment of full and super levels): He\one\ (40,51), He\two\ (13,30), C\one\ (14,26), C\two\ (14,26), C\three\ (62,112), N\one\ (44,104), N\two\ (23,41), N\three\ (25,53), O\one\ (19,51), O\two\ (30,111), O\three\ (50,86), Ne\one\ (70,139), Ne\two\ (22,91), Ne\three\ (23,71), Na\one\ (22,71), Mg\two\ (22,65), Mg\three\ (31,99), Al\two\ (26,44), Al\three\ (17,45), Si\two\ (31,59), Si\three\ (33,61), Si\four\ (37,48), S\two\ (56,324), S\three\ (48,98), S\four\ (27,67), Ar\one\ (56,110), Ar\two\ (134,415), Ar\three\ (32,346), Ca\two\ (21,77), Ca\three\ (16,40), Ca\four\ (18,69), Sc\two\ (38,85), Sc\three\ (25,45), Ti\two\ (37,152), Ti\three\ (33,206), V\one\ (1,1), Cr\two\ (28,196), Cr\three\ (30,145), Cr\four\ (29,234), Mn\two\ (25,97), Mn\three\ (30,175), Fe\one\ (44,136), Fe\two\ (275,827), Fe\three\ (83,698), Fe\four\ (100,1000), Fe\five\ (47,191), Co\two\ (136,2747), Co\three\ (124,3917), Co\four\ (37,314), Co\five\ (32,387), Ni\two\ (59,1000), Ni\three\ (47,1000), Ni\four\ (36,200), and Ni\five\ (46,183).
We note that the atomic data for our \cmfgen\ calculations differs from our Sedona calculations.

We start the \cmfgen\ simulations for each wedge at 10--15\,d after explosion and typically a few days before $B$-band maximum for that given wedge (as computed by Sedona).
For the first time step in the sequence, we compute the model with the temperature fixed and assuming steady-state. Once this initial model is converged, we evolve the model with time dependence using a time increment equal to 10\,\% of the current time until we reach about 30-35\,d.
The goal is to cover from $B$-band maximum until 15\,d later in the \cmfgen\ sequence (the time of $B$-band  maximum computed by Sedona and \cmfgen\ differ).

For each wedge and time, we initially compute the $\gamma$-ray energy deposition profile using a pure-absorption, grey transport solver with $\gamma$-ray opacity equal to 0.06\,$Y_{\rm e}$\,cm$^2$\,g$^{-1}$, where $Y_{\rm e}$ is the material electron fraction.
Once this model is converged, we rerun it by scaling the depth-dependent energy-deposition profile so that the volume integrated decay power (i.e. outgoing luminosity) equals the 2D outgoing luminosity computed by Sedona for that wedge direction.
With this approach, we mimic in 1D the luminosity of the 2D model while capturing the critical non-LTE effects inherent to the radiative transfer.
Similar to our 1D LTE results, we likewise refer to these calculations and observables as ``1D'' non-LTE, despite their multidimensional considerations.

\subsection{Construction of Pseudo-2D non-LTE Observables}
\label{subsec:pseudo2d}

In this work, we aim to simultaneously analyze non-LTE and multidimensional contributions to the synthetic observables.
As proper multidimensional, non-LTE calculations are generally unfeasible at this time, we concatenate our three sets of radiative transfer results to produce an approximated facsimile of such a calculation.
We refer to this new set of synthetic observables as ``pseudo-2D non-LTE''.
Specifically, we take the differences in flux at each wavelength between the 1D and 2D LTE calculations (the \tdltecolor{} bands in Figure \ref{fig:1d_vs_2d_lte_spectra}, i.e.\ the multidimensional correction) and apply them to the 1D non-LTE spectra at each time step for each line of sight.
Likewise, for the theoretical photometry of our models, we take the flux difference between the 1D LTE and 2D LTE in each band and apply them to the 1D non-LTE to construct our pseudo-2D non-LTE light curves.
The total energy of these pseudo-2D non-LTE observables are conserved under this strategy (i.e.\ the bolometric light curves between all four sets of synthetic observables are equivalent within a negligible tolerance).
The 2D LTE calculation has a time step of 0.5 d, while the 1D LTE and non-LTE calculations have a time step of 1.5 d, in sync with every third 2D LTE time.

The utility of the pseudo-2D non-LTE synthetic observables is marginally limited in that their construction is not physically motivated.
Additionally, there may be non-LTE processes that are exclusively relevant in multiple dimensions.
However, these corrected observables are useful for identifying regions of spectra and times where multidimensional features may not be captured by the 1D non-LTE calculations.
We include these pseudo-2D non-LTE synthetic observables in comparisons with observed SNe Ia spectra, light curves, and population correlations in following sections.

\section{Synthetic Observables}
\label{sec:results}


\begin{table*}
\centering
\caption{Wedge Details}
\begin{tabular}{c|c|c|c|c|c|c}
Wedge & v$_{IGE}$\footnote{Representative velocity for the outer edge of the IGE-rich region, arbitrarily chosen where $X_{Ni} < 0.1$} & L$_{bol,max}$ & \multicolumn{4}{c}{M$_{B,max}$} \\ \hline
& $10^9$ cm s$^{-1}$ & $10^{42}$ erg s$^{-1}$ & \multicolumn{4}{c}{\textit{mag}}\\ 
 & & & \multicolumn{2}{c|}{LTE} & \multicolumn{2}{c}{non-LTE} \\
 &  &  & 2D & 1D & 1D & pseudo-2D \\ \hline \hline
1 & 1.49 & 8.33 & -18.94 & -18.86 & -18.77 & -18.87 \\
2 & 1.42 & 9.20 & -19.06 & -19.06 & -18.81 & -18.81 \\
3 & 1.37 & 9.81 & -19.13 & -19.13 & -18.87 & -18.88 \\
4 & 1.30 & 10.36 & -19.18 & -19.16 & -18.91 & -18.95 \\
5 & 1.25 & 10.85 & -19.22 & -19.18 & -18.94 & -19.00 \\
6 & 1.20 & 11.36 & -19.25 & -19.19 & -18.98 & -19.05 \\
7 & 1.16 & 11.64 & -19.27 & -19.19 & -19.01 & -19.10 \\
8 & 1.12 & 12.13 & -19.29 & -19.19 & -19.03 & -19.14 \\
9 & 1.11 & 12.49 & -19.31 & -19.20 & -19.06 & -19.17 \\
10 & 1.09 & 12.67 & -19.32 & -19.22 & -19.08 & -19.18 \\
11 & 1.01 & 12.93 & -19.33 & -19.20 & -19.09 & -19.22 \\
12 & 1.05 & 13.15 & -19.33 & -19.19 & -19.09 & -19.23 \\
13 & 1.17 & 13.37 & -19.33 & -19.28 & -19.15 & -19.20 \\
14 & 1.19 & 13.70 & -19.33 & -19.28 & -19.15 & -19.20 \\
15 & 1.17 & 13.85 & -19.32 & -19.35 & -19.17 & -19.14 \\
\end{tabular}
\label{table:wedges}
\end{table*}

We summarize the details of our wedge profiles and calculations in Table \ref{table:wedges}.
The diversity of these wedges are demonstrated by their varying values of $v_{IGE}$, which roughly represents the velocity extent of IGE-group material, and peak bolometric luminosities.
The bolometric luminosity for a given wedge is the same between our calculations as our method prescribes.

In the follow subsections, we show the results of our various radiative transfer calculations.
In Section \ref{subsec:1d_vs_2d_lte} we compare the 1D and 2D LTE synthetic spectra to reveal the multidimensional differences between the two calculations.
In Section \ref{subsec:lte_vs_nlte}, we show the differences in synthetic spectra between the 1D LTE and 1D non-LTE calculations to determine the impact of higher accuracy physics for this model.
Finally, we compare the light curves for our sets of observables in \ref{subsec:model_lightcurves}.

\subsection{1D vs. 2D LTE}
\label{subsec:1d_vs_2d_lte}

\begin{figure*}
    \centering    
    \includegraphics[width=1.0\textwidth]{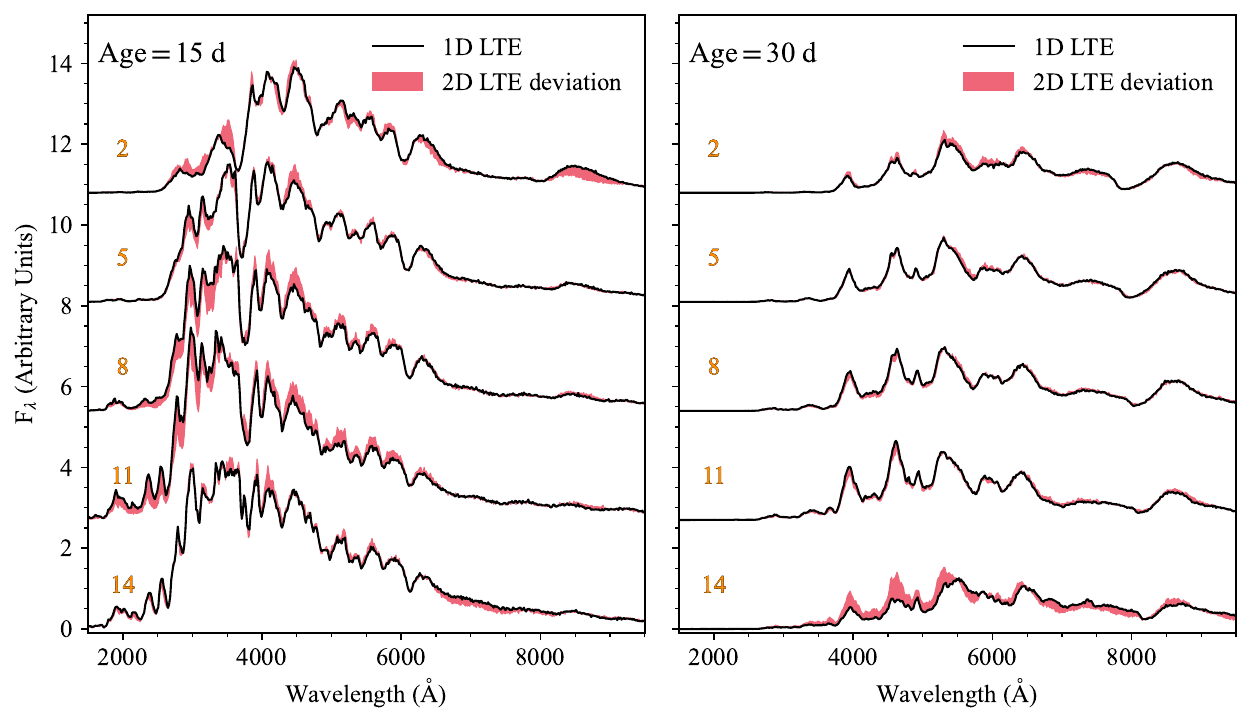}
    \caption{LTE spectra at a selection of lines of sight at 15 and 30 days after the explosion.
    Spectra from our 1D LTE calculations are shown in black while the \tdltecolor{} shading indicates how the 2D LTE spectra deviate from the 1D LTE spectra (i.e.\ the outer edge of the \tdltecolor{}  shading is the 2D spectra).
    The lines of sight are labeled in orange, corresponding with the wedges in Figure \ref{fig:boos21_ejecta}.
    }
    \label{fig:1d_vs_2d_lte_spectra}
\end{figure*}

We show in Figure \ref{fig:1d_vs_2d_lte_spectra} the difference in synthetic spectra between our 1D wedge and 2D full ejecta calculations in LTE at a selection of lines of sight at 15 days and 30 days after the time of explosion, the former of which is around the time of maximum $B$-band magnitude (see Figure \ref{fig:appendix_lte} in the Appendix for all lines of sight).
The 2D deviation from the 1D counterpart, illustrated in \tdltecolor{}, is most significant at early times in addition to being dependent on line of sight.
For example, the equatorial line of sight shows modest differences with a slightly redder spectrum in the 2D calculation in the near UV and visual at 15 days post-explosion.
This color disagreement persists at southern lines of sight, including at wedge 11 where it is most significant, but weakens and reverses at northern lines of sight.
Wedge 11 also sees the most notable change in the trough shape of its Ca \textsc{ii} H\&K absorption feature ($\sim3800$ \AA).
Additionally, there is a modest excess of flux redward of $6000$ \AA\ for the 2D spectra at the southernmost lines of sight, while the northernmost lines of sight see a fairly significant lack of flux in this region, including the three northernmost viewing angles which show a considerable departure in the Ca \textsc{ii} triplet at $8500$ \AA.
The average absolute difference in flux between the 1D and 2D spectra as a percentage of the total flux at 15 days is 11\%, with a median of 10\%.

At 30 days post-explosion, the differences in spectra between the 1D and 2D calculations are smaller in flux.
The 1D and 2D spectra show remarkably few differences at most lines of sight.
For our boundary wedges (1 and 15), we find non-trivial differences between the 1D and 2D LTE spectra at later times, which may be attributable to spurious symmetry effects in the hydronuclear and/or radiative transfer calculations.
Wedges 13 and 14 also show notable deviations.
This is likely due to the three southernmost wedges containing significantly more $^{56}$Ni than the other wedges, as a result of the $^{56}$Ni distribution from this typical double detonation model being centered in the southern hemisphere.
We find an average and median absolute difference in flux between the 1D and 2D spectra as a percentage of the total flux at 30 days of 14\% and 9\%, respectively.


\subsection{1D LTE vs. 1D non-LTE}
\label{subsec:lte_vs_nlte}

\begin{figure*}
    \centering    
    \includegraphics[width=1.0\textwidth]{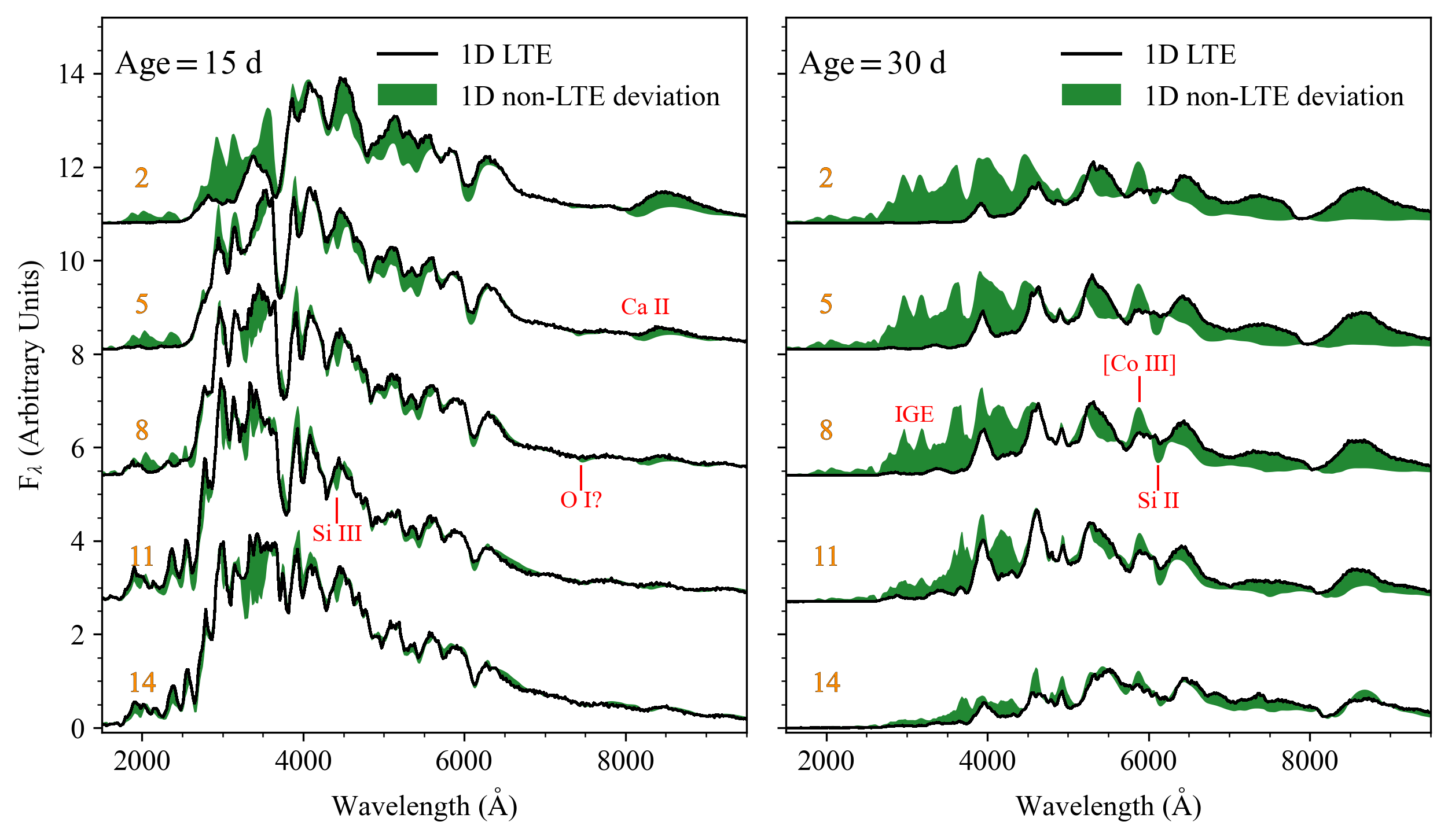}
    \caption{1D spectra at a selection of lines of sight at 15 and 30 days after the explosion.
    Spectra from our 1D LTE calculations are shown in black while the \odnltecolor{} shading indicates how the 1D non-LTE spectra deviate from the 1D LTE spectra (i.e.\ the outer edge of the \odnltecolor{} shading is the 1D non-LTE spectra).
    The lines of sight are labeled in orange, corresponding with the wedges in Figure \ref{fig:boos21_ejecta}.
    Notable spectral features are labeled in red.
    }
    \label{fig:nlte_vs_lte_spectra}
\end{figure*}

In Figure \ref{fig:nlte_vs_lte_spectra}, we again show 1D LTE spectra for several lines of sight at 15 and 30 days after explosion, but now with the differences with the 1D non-LTE spectra, illustrated in \odnltecolor{}.
Importantly, the difference in 1D LTE and non-LTE spectra are, for the most part, greater than that between 1D and 2D LTE calculations.
In other words, non-LTE effects are of greater consequence on the observables than those that arise from multidimensionality.

At 15 days post-explosion, a significant discrepancy between the non-LTE and LTE spectra is in the UV.
Northern wedges show higher flux blueward of the Ca \textsc{ii} H\&K absorption feature and particularly at wedge 2 where the emission is considerably larger for the non-LTE spectra.
A notable aspect of this disagreement is the pair of adjacent peaks centered at $\sim3000$ \AA\,, which is largely formed due to IGE line blanketing (see \citealt{derkacy_2020}), most easily observed in the non-LTE spectrum of wedge 2.
This distinct feature also exists in the LTE spectra between wedges 5 and 8 and is less prominent at more southern lines of sight in both non-LTE and LTE.
However, it is prominently produced in the non-LTE calculation for wedge 2 where it does not exist in the LTE spectra.
Conversely, at equatorial and southern lines of sight, the non-LTE spectra show lower emission in the near-UV around $3500$ \AA\ compared to the LTE spectra.

Overall, the emission in the visual region of the spectra at 15 days post-explosion is lower in the non-LTE calculation, with the strength of the variation decreasing from the northernmost line of sight.
The discrepancy between the two sets of spectra in this region fluctuates somewhat across wavelengths.
For example, there appears to be fairly consistent agreement of the flux in the Fe \textsc{ii} absorption complex at $\sim5000$ \AA.
Redward of this, the non-LTE spectra is similarly featured but suppressed, especially at northern and equatorial lines of sight.
Blueward of the shared absorption feature at $\sim5000$ \AA, the non-LTE spectra produce a prominent absorption feature at $\sim4400$ \AA\ that is not present in any of the LTE spectra.
We attribute this to Si \textsc{iii} $\lambda 4560$, which appears in some observed SNe Ia (see Section \ref{subsec:obs_spectral_comparisons}).
The non-LTE spectra also show higher UV emission, especially at northern and equatorial lines of sight, in addition to slightly higher peaks on either side of the Si \textsc{ii} $\lambda4130$ absorption feature at $\sim4000$ \AA.
Additionally, the non-LTE spectra show a deeper Si \textsc{ii} $\lambda6355$ absorption feature and an altered near-IR Ca \textsc{ii} triplet at northern and equatorial lines of sight. 
Lastly, we observe a very slight absorption feature at $\sim7500$ \AA\ at some lines of sight in the non-LTE spectra, which may be O \textsc{i} \citep{Zhao_2016}.
The integrated absolute differences in flux between the 1D LTE and non-LTE spectra as a percentage of the total flux have an average and mean of 16\% and 11\%, respectively, at 15 days post-explosion.

The difference in spectra between non-LTE and LTE calculations grows much greater as time increases beyond the maximum light, exemplified in the right panel of Figure \ref{fig:nlte_vs_lte_spectra} where the spectra are compared at 30 days post-explosion, and declines in strength from the northernmost wedge.
The two sets of spectra are disjoint at most lines of sight across all wavelengths except in the Fe \textsc{ii}-dominated region around $5000$ \AA.
The non-LTE spectra are significantly bluer than their LTE counterparts, with considerable excesses in UV and B flux and a flattened near-IR in all but the two southernmost lines of sight.
Similar stark differences between \cmfgen\ and Sedona spectra around this phase have previously been shown in \citet{blondin_2022}.

There are exclusive, prominent features in the non-LTE spectra at this time, including the pair of peaks centered at $\sim3000$ \AA, which likely appears due to increasing IGE line blanketing in combination with a stronger UV continuum in the non-LTE calculation.
Ca \textsc{ii} H\&K and Si \textsc{ii} $\lambda$6355 can also be observed to persist through 30 days exclusively in the non-LTE spectra for many lines of sight.
The absorption feature around 4400 \AA\ is also unique to the non-LTE spectra and may be connected to the Mg \text{ii}-dominant absorption feature in the same region in maximum light spectra.
Lastly, there is a strong peak just blueward of Si \textsc{ii} $\lambda$6355 in the non-LTE spectra where the LTE spectra are relatively flat.
This is very likely due to forbidden line transition [Co \textsc{iii}] $\lambda$5888 \citep{dessart_2014_coIII}.

The average and mean integrated absolute difference in flux between these two calculations as a percentage of the total flux are 69\% and 74\%, respectively, at 30 days post-explosion.

\begin{figure*}
    \centering    
    \includegraphics[width=1.0\textwidth]{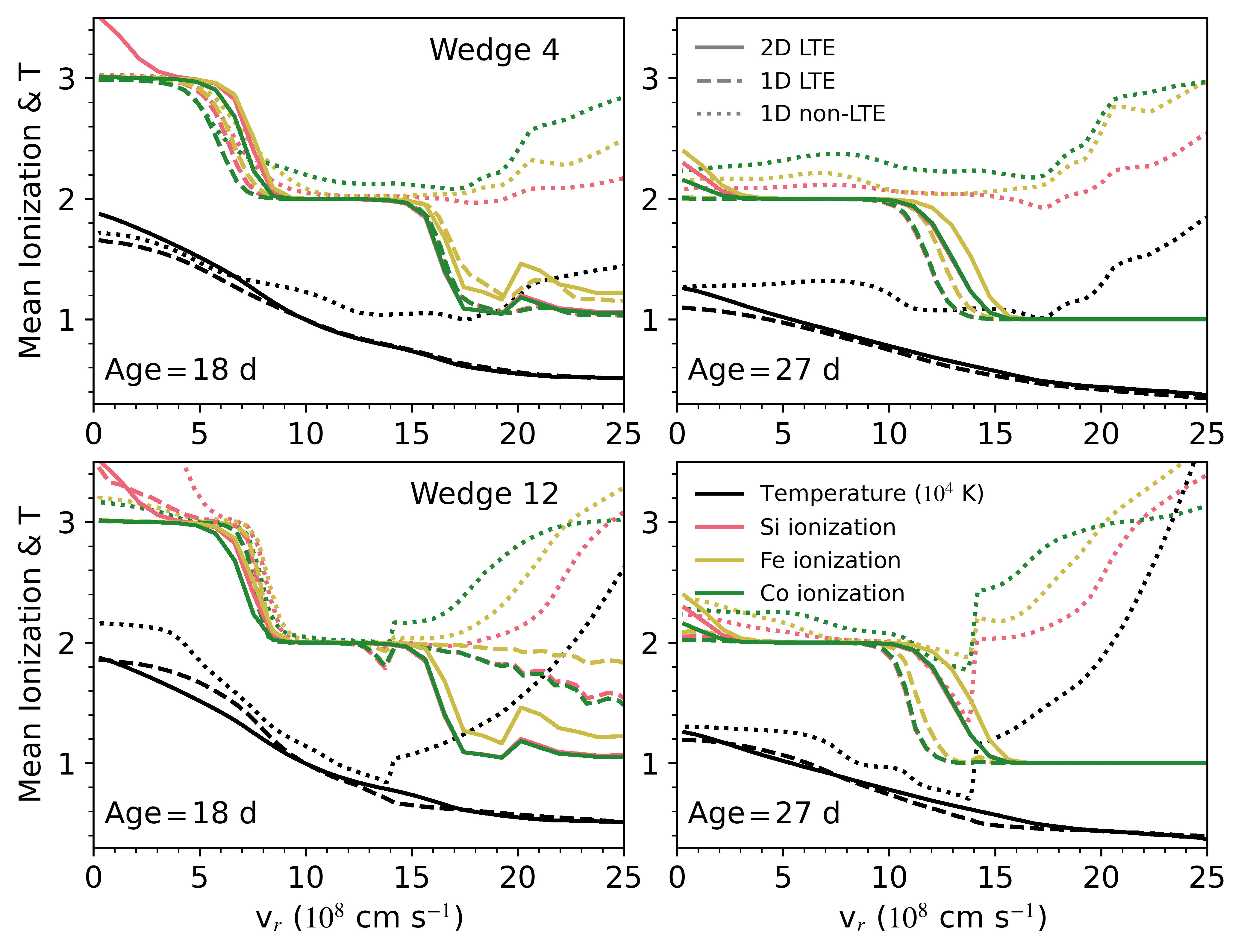}
    \caption{Temperature and ionization profiles for our three radiative transfer calculations for wedges 4 and 12 at 18 and 27 days post-explosion.
    The 2D LTE, 1D LTE, and 1D non-LTE calculations are shown with solid, dashed, and dotted lines, respectively.
    The temperature and Si, Fe, and Co mean ionization values are displayed in black, red, yellow, and green, respectively.
    }
    \label{fig:ionization_profiles}
\end{figure*}

To elucidate the origin of the differences between the non-LTE and LTE spectra, we show in Figure \ref{fig:ionization_profiles} the temperature and ionization profiles for both of these treatments for wedges 4 and 12 near maximum light and nine days later.
At low velocities ($\lesssim10,000$ km s$^{-1}$) beneath the photosphere for both wedges at maximum light (left panels of Figure \ref{fig:ionization_profiles}), the temperature and ionization states are similar between the different calculations, indicating that LTE conditions are generally met within this region.
There is some disparity between the Si ionization states at very low velocities around maximum light. 
However, this region is predominately composed of IGE, and thus the Si ionization at these depths is inconsequential.

The maximum light similarities in ejecta states for these wedges extends out to $\sim15,000$ km s$^{-1}$, outward of which the temperature and ionization diverge between the non-LTE and LTE calculations.
This point of divergence is associated with the density discontinuity of the double detonation model, arising from the interface between the core and shell detonations.
At these higher velocities where the shell detonation ashes are found, the non-LTE effects become dominant due to the optically-thin material, leading to much higher ionization.
Additionally, a spurious spike in mean ionization can be observed in this density discontinuity region in the 1D LTE ionization profile of wedge 12, which may arise due to the 1D ejecta profile improperly encapsulating the density discontinuity, LTE issues, or a combination of both.

The general agreement between ionization states at low velocities is weakened by 27 days (right panels of Figure \ref{fig:ionization_profiles}), with the non-LTE profiles showing moderately higher ionization, along with temperature profiles that have a much shallower decline than the LTE counterparts.
This is largely due to the significant reduction in density throughout the ejecta by this time.
At high velocities, the state of the ejecta is even more starkly different between non-LTE and LTE at this later time, with a notable increase in temperature and mean ionization state at high velocities in the non-LTE calculation, especially for wedge 12.

\citet{blondin_2022} show a similar comparison for a 1D SN Ia toy model in their Figure 7, though we note an appreciable difference in input model as well as plotted times.
\citet{blondin_2022} find temperature and Co ionization profiles that similarly differ between Sedona and \cmfgen\ calculations at low velocities around maximum light.
However, the comparison from \citet{blondin_2022} shows \cmfgen\ temperature and ionization profiles that decline monotonically around maximum light, which is likely attributable to their smooth and monotonic toy model, as opposed to our non-LTE profiles which see an increase at high velocities.
At later times, the difference in profiles of our work more closely resembles that of \citet{blondin_2022}.



In summary, we emphasize the significance of the departure from LTE spectra under a non-LTE treatment of the radiative transfer shown in Figure \ref{fig:nlte_vs_lte_spectra}.
These spectral differences are a consequence of the additional physics of the non-LTE calculations, which leads to generally higher temperatures, more ionization, and bluer spectra overall.
While the spectra at 15 days after explosion ($\sim$maximum light) are somewhat globally similar, there is a non-trivial shift in the spectral shape in addition to the previously described exclusive Si \textsc{iii} $\lambda 4560$ line.
More significantly, the 30 day post-explosion spectra show notable differences between non-LTE and LTE, with very little agreement across the majority of the optical spectrum for most lines of sight.
Thus, the use of non-LTE becomes necessary for quantitative predictions after the time of maximum light.

\subsection{Light Curves}
\label{subsec:model_lightcurves}

\begin{figure*}
    \centering
    \includegraphics[width=1.0\textwidth]{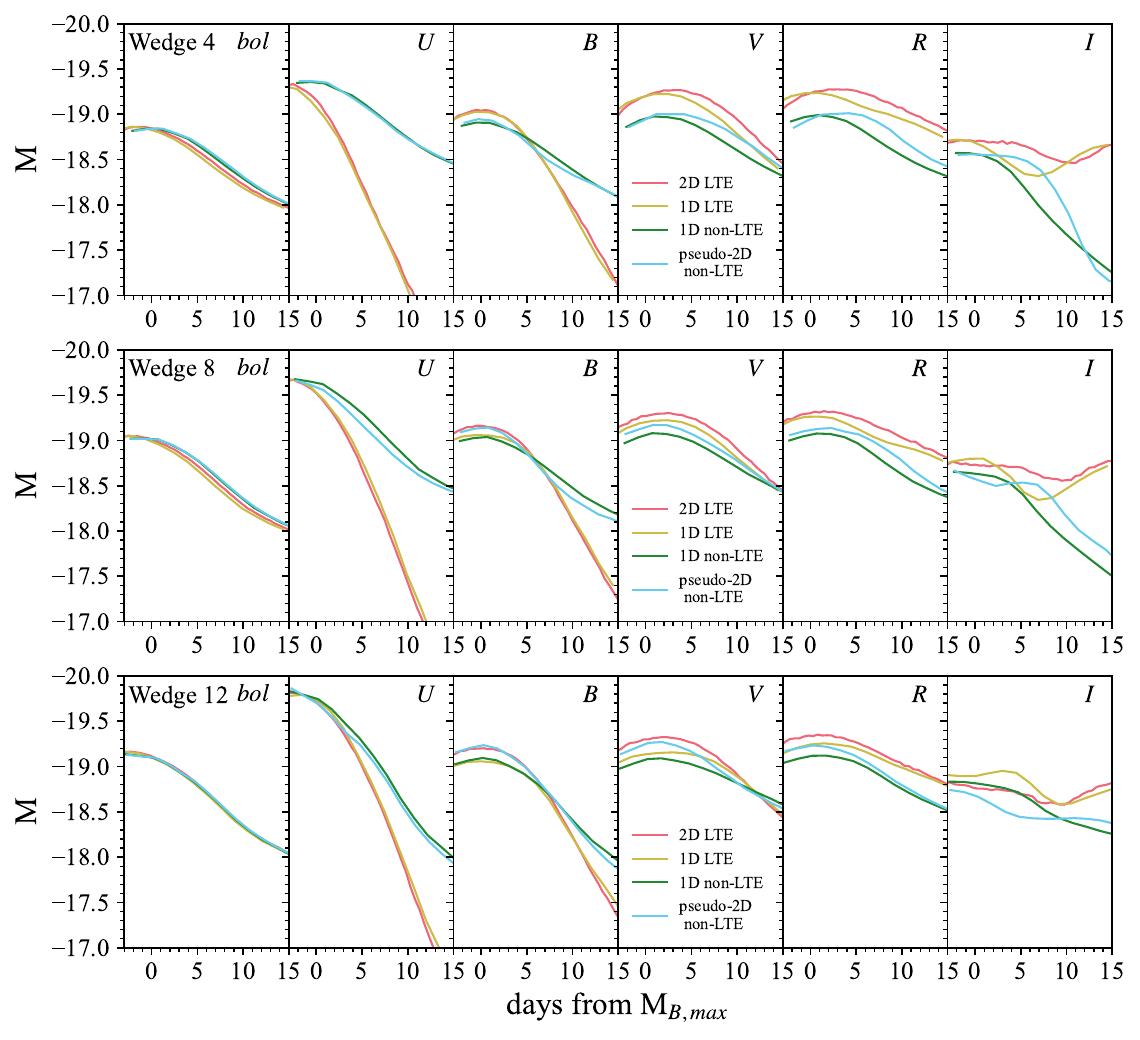}
    \caption{Bolometric and UBVRI light curves for our four sets of synthetic observables at three different lines of sight, adjusted to their respective $B$-band maxima times.
    }
    \label{fig:model_light_curves}
\end{figure*}

We show in Figure \ref{fig:model_light_curves} the light curves for our four sets of synthetic observables for a representative selection of viewing angles.
The bolometric light curves for a given line of sight are effectively identical due to our imposed constraint on the non-LTE and 1D LTE calculations that they match the bolometric output from the 2D calculations.
However, some bolometric light curves appear slightly offset in Figure \ref{fig:model_light_curves} due to small shifts in the time of maximum light in the $B$-band between the sets of observables.
The change in time of $B$-band peak is no more than 1.8 d for all lines of sight, with an average difference of 1.0 d between the 1D LTE and 1D non-LTE light curves, the largest shift between any corresponding set of models.

The multidimensional effect in LTE on the light curves can be determined by comparing the \odltecolor{} and \tdltecolor{} lines in Figure \ref{fig:model_light_curves}.
This effect is relatively minor in bands $U$, $B$, $V$, and $R$, where the light curve shapes between the 1D and 2D LTE are very similar, with slight disagreement in peak magnitudes in $B$ and $V$ for some wedges.
The 1D calculation does a much poorer job of reproducing the $I$-band, however, as the number and location of the local extrema vary non-trivially between 1D and 2D LTE.

As indicated by the differences between the spectra in Figure \ref{fig:nlte_vs_lte_spectra}, there is a dramatic change between the LTE and non-LTE light curves, especially relative to the aforementioned minor differences from dimensionality.
While the $U$-band peaks at a similar magnitude between LTE and non-LTE treatments, the declines see significantly different shapes and rates, especially at northern latitudes (e.g.\ the top panel of Figure \ref{fig:model_light_curves}).
The peak $B$-band magnitudes vary somewhat between the non-LTE and LTE, depending on the line of sight.
Crucially, the shape of the $B$-band light curve is starkly different in the non-LTE calculation with a consistently shallower decline from peak to 15 days post.
We discuss this change in $B$-band light curve, including its relevance to the Phillips relation in particular, below in Section \ref{subsec:obs_correlations}.

In longer wavelength bands, we observe lower emission in the non-LTE light curves compared to LTE for most times between 0 and 15 days after maximum light. 
For the $V$- and $R$-band, this difference is heavily dependent on line of sight, as shown in Figure \ref{fig:model_light_curves}, where the peaks have a larger disparity at northern lines of sight.
The shapes of the light curves also vary between the radiative transfer calculations, especially in $V$-band where the decline is shallower in non-LTE.
Additionally, we see a significant change in the post-peak $I$-band light curves, with the non-LTE exhibiting a mostly monotonic decline compared to the secondary local maxima seen in the LTE.
This lack of secondary maximum in the non-LTE $I$-band light curves may be attributable to the stunted recombination of IGE elements in these calculations (see Figure \ref{fig:ionization_profiles})  \citep{kasen_2006_Iband}.
We note the relatively large discrepancy between the 1D and pseudo-2D non-LTE in the $I$-band, corresponding to the disparity between the 1D and 2D LTE light curve in this band, which indicates this band is somewhat questionable in its interpretation.
Future work will strive to reduce uncertainties in the $I$-band to better understand how both multidimensionality and non-LTE affect local extrema.


\section{Comparison to Observation}
\label{sec:compare_to_obs}

In this section, we compare our synthetic observables to observed properties of SNe Ia.
We compare our models to individual events and correlations of SNe Ia as a population in Sections \ref{subsec:obs_spectral_comparisons} and \ref{subsec:obs_correlations}, respectively.

\subsection{Comparison to Individual SNe Ia}
\label{subsec:obs_spectral_comparisons}

Here we compare our synthetic observables from a pair of observed SNe Ia with spectra and light curves from our LTE and non-LTE calculations.
To find spectral matches between our calculations and observations, an attempt was made to use SNID \citep{blondin_tonry_2007}, which is intended to be used with observed spectra of SNe to classify them and estimate their redshift and epoch based on a set of templates.
As the SNID algorithm removes the pseudo-continuum prior to fitting, however, we found that excellent matches determined and displayed by SNID were relatively poor matches when plotting their actual spectra due to differing spectral shapes between observation and our models.
We believe that a tool that could effectively match synthetic and observed spectra while still considering for spectral shape would be beneficial to the field, especially as synthetic spectra from a variety of models improve in accuracy and the number of observed events increases.

An extensive search to determine the best matches of observables between our calculations and observed events was not undertaken due to the limited theoretical data set and scope of this work, in addition to the lack of efficient event-matching tools that the authors are aware of.
Rather, we selected two samples of SNe Ia (\citealt{hicken_2009} and \citealt{burrow_2020}), identified events therein that had similar $B$-band peak magnitudes and declines rates to our non-LTE observables, and inspected the spectra by eye to establish ``matches''.
We additionally compare to SN 2011fe.
In the following subsections, we compare the observables from these normal SNe Ia with our wedges that are most similar.

\begin{table*}
\caption{Observational comparisons}
\begin{tabular}{c|c|c|c|c|c|c|c}
Event & z & E(B-V)$_{host}$& Wedge & \multicolumn{2}{c|}{M$_{B,max}$} & \multicolumn{2}{c}{$\Delta$M$_{15,B}$} \\ \hline 
& & \textit{mag} & &\multicolumn{2}{c|}{\textit{mag}} & \multicolumn{2}{c}{\textit{mag}}  \\ 
 & & & & Observed & Model\tablenotemark{a} & Observed & Model\tablenotemark{a} \\ \hline \hline
SN 2011fe & 0.001 & -- & 12 & $-19.21\pm0.15$\tablenotemark{b} & $-19.23$ & $1.21\pm 0.03$\tablenotemark{b} & 1.37 \\
SN 1999ek & 0.018 & 0.184 & 11 & $-19.21 \pm0.47$\tablenotemark{c} & $-19.22$ & $1.21\pm0.37$\tablenotemark{c} & 1.22 \\
SN 2007bd & 0.032 & 0.025 & 7 & $-19.28\pm0.10$\tablenotemark{d} & $-19.10$ & $1.24\pm 0.01$\tablenotemark{e} & 1.02 \\
\end{tabular}
\tablenotetext{a}{Pseudo-2D non-LTE}
\tablenotetext{b}{\citealt{Richmond_2012}}
\tablenotetext{c}{\citealt{hicken_2009}}
\tablenotetext{d}{\citealt{burrow_2020}}
\tablenotetext{e}{\citealt{stritzinger_2011}}
\label{table:observed}
\end{table*}

Spectra of observed events are modified correcting for distance and redshift, as well as Milky Way and host galaxy reddening, using the SNooPy Python tool for SNe Ia \citep{Burns_2011} to deredden.
Light curves of observed events are modified by correcting for distance, time dilation, galactic extinction and host galaxy extinction (we do not consider host galaxy extinction for SN 2011fe).
Values for galactic extinction are drawn from \citet{schlafly_2011} while host galaxy extinction is determined using $R_V=1.7$ with color excesses given by \citet{hicken_2009}.
We show a summary of these observations and our corresponding model wedges in Table \ref{table:observed}.
Data for our model in this table is given for only the pseudo-2D synthetic observables.

\subsubsection{SN 2011fe}

\begin{figure}
    \centering    
    \includegraphics[width=0.5\textwidth]{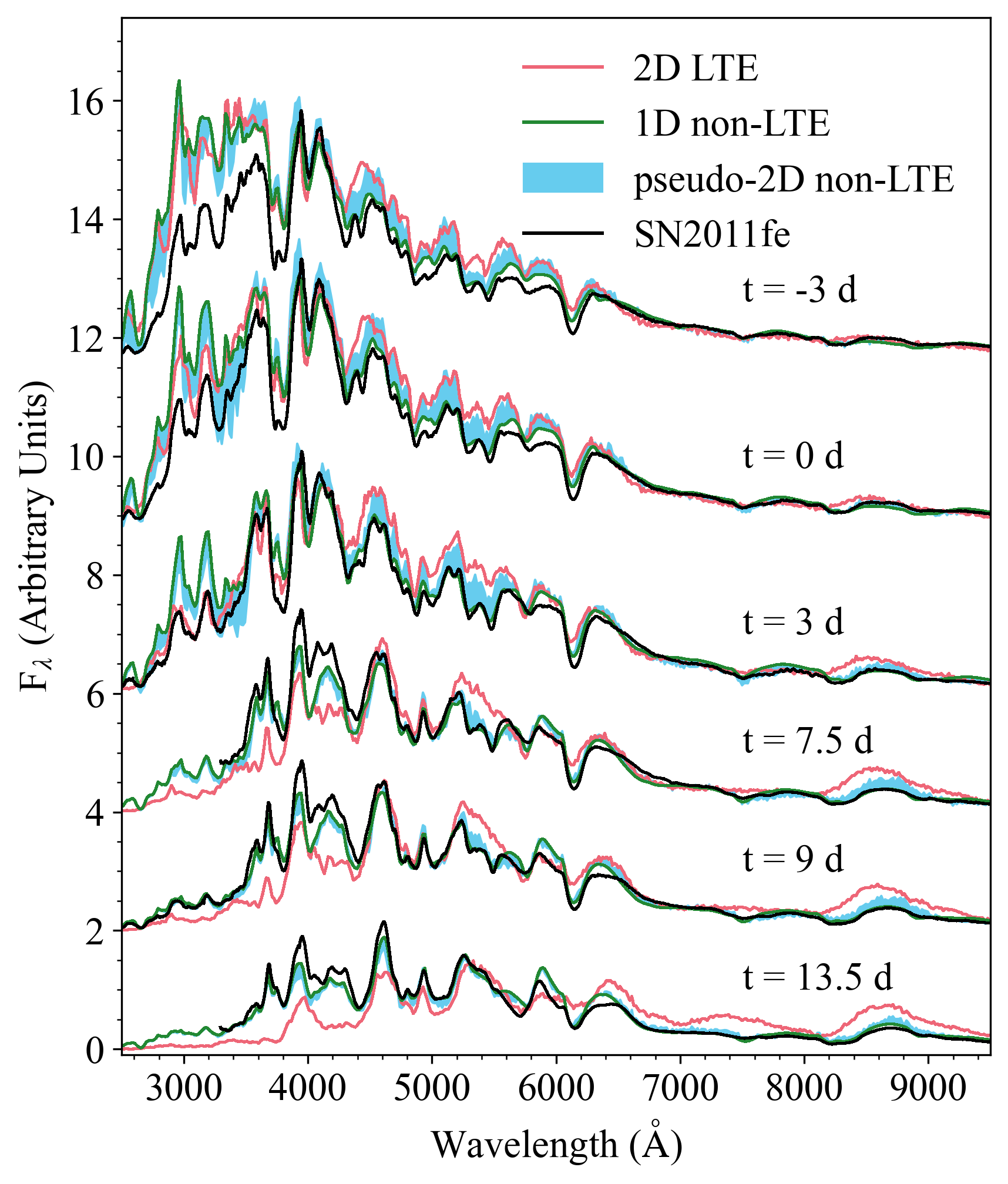}
    \caption{Spectrum of SN 2011fe \citep{pereira_2013,mazzali_2014} compared with our 1D non-LTE and 2D LTE spectra for wedge 12 of our model.
    The \tdnltecolor{} shading indicates the difference between 1D non-LTE and pseudo-2D non-LTE spectra at the given time.
    Spectra are labeled based on the time relative to their respective $B$-band maximum.
    }
    \label{fig:2011fe_spectra}
\end{figure}

Spectra of SN 2011fe \citep{pereira_2013,mazzali_2014} are compared with our 2D LTE and 1D non-LTE results for wedge 12 of our model in Figure \ref{fig:2011fe_spectra}.
We additionally show our pseudo-2D non-LTE spectra (\tdnltecolor{}) as a deviation from the the 1D non-LTE spectra, which gives an indication of where multidimensional effects may apply.
We emphasize that this deviation shown in Figure \ref{fig:2011fe_spectra}, and the following spectral comparisons, display the differences between the 1D non-LTE and pseudo-2D non-LTE results with consideration for a change of the time of $B$-band maximum between sets of synthetic observables (see Figure \ref{fig:1d_vs_2d_lte_spectra} for an illustration of the multidimensional spectral differences at the same post-explosion times).

Before and at maximum light, SN 2011fe is well-reproduced by the 1D non-LTE spectra at wavelengths blueward of the Ca \textsc{ii} H\&K feature, including the Si \textsc{iii} $\lambda 4560$ absorption feature.
In this same region, the 2D LTE spectra fails to reproduce some features of SN 2011fe, in addition to having modestly higher emission in the optical.
Much of the pseudo-2D non-LTE spectra deviates from the 1D non-LTE here, however, suggesting that the strong agreement between SN 2011fe and the 1D non-LTE spectra may be driven by the reduced multidimensionality of the calculation.
Additionally, the UV spectra of SN 2011fe is not well-replicated by any of our synthetic observables at maximum light.

By 7.5 days, the disparity between the LTE and non-LTE agreement with SN 2011fe is even greater, in addition to the multidimensional correction being lower.
At this and later times, non-LTE spectra more accurately reproduce the near-UV features of SN 2011fe, in addition to the waning Ca \textsc{ii} H\&K feature.
The non-LTE spectra shows improved agreement in the optical, with a lack of suppression at $\sim4250$\AA\ and a S \textsc{ii} $\lambda5640$ absorption feature that persists longer beyond maximum light.
The very near IR spectra is also much improved in the non-LTE calculations, including the Ca \textsc{ii} triplet at $\sim8500$\AA.

\begin{figure}
    \centering    
    \includegraphics[width=.50\textwidth]{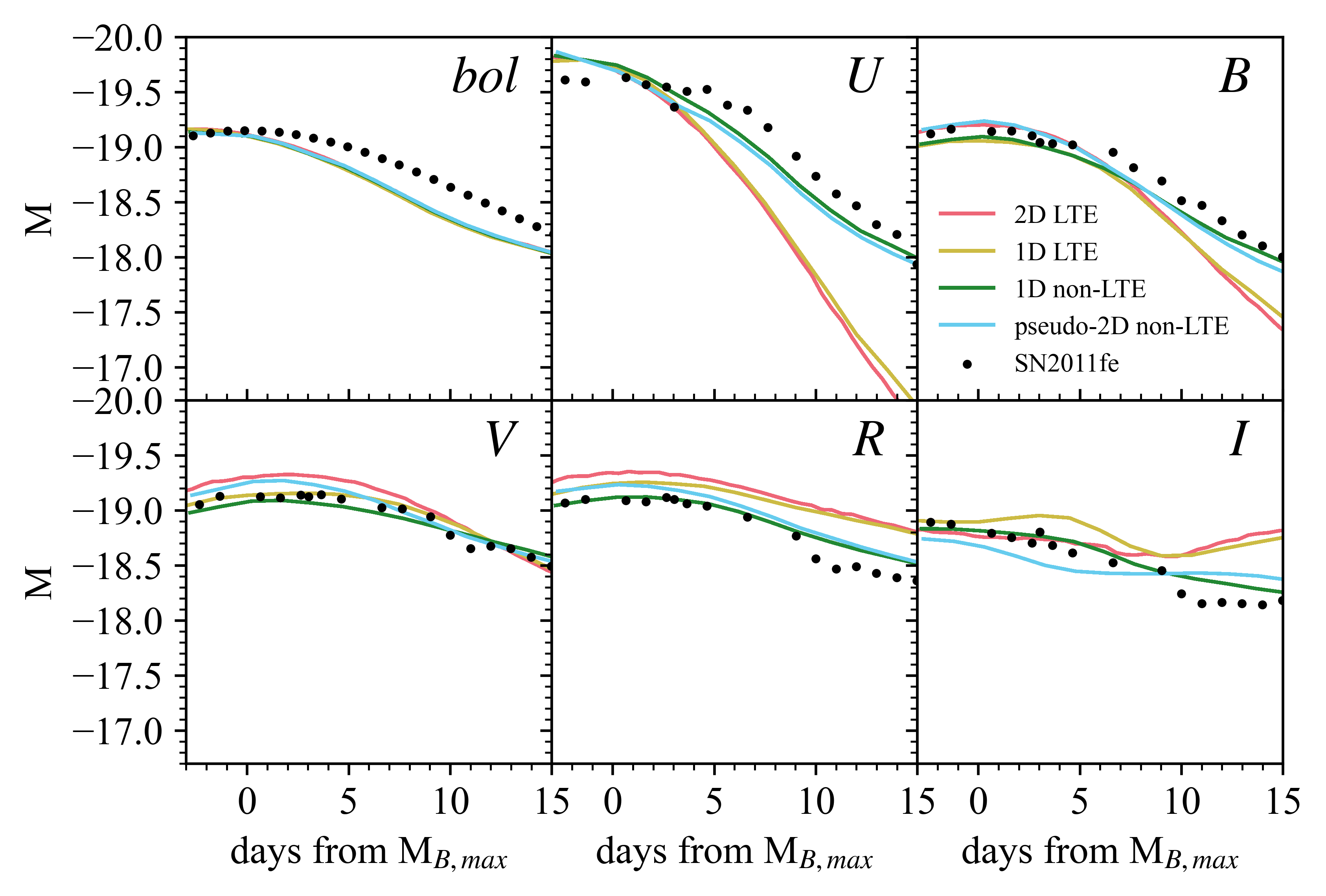}
    \caption{Light curves of SN 2011fe \citep{munari_2013,tsvetkov_2013} compared with that from each of our radiative transfer calculations of wedge 12 of our model.
    }
    \label{fig:2011fe_lc}
\end{figure}

We compare the photometry of SN 2011fe \citep{munari_2013,tsvetkov_2013} with each set of our synthetic observables for wedge 12 in Figure \ref{fig:2011fe_lc}.
While the shape of the pseudo-bolometric light curve of SN 2011fe is not well fit by our model constraint, the non-LTE $UBVRI$ light curves show a strong agreement with observation.
Most notably, the 1D non-LTE $U$- and $B$-band are remarkably consistent with that from SN 2011fe, which is a significant departure from the LTE result.
There is a modest multidimensional correction for the $B$-band at max light for this wedge, however.
The $V$- and $R$-bands also show moderately improved agreement with SN 2011fe under non-LTE.

\subsubsection{SN 2007bd}
\begin{figure}
    \centering    
    \includegraphics[width=0.5\textwidth]{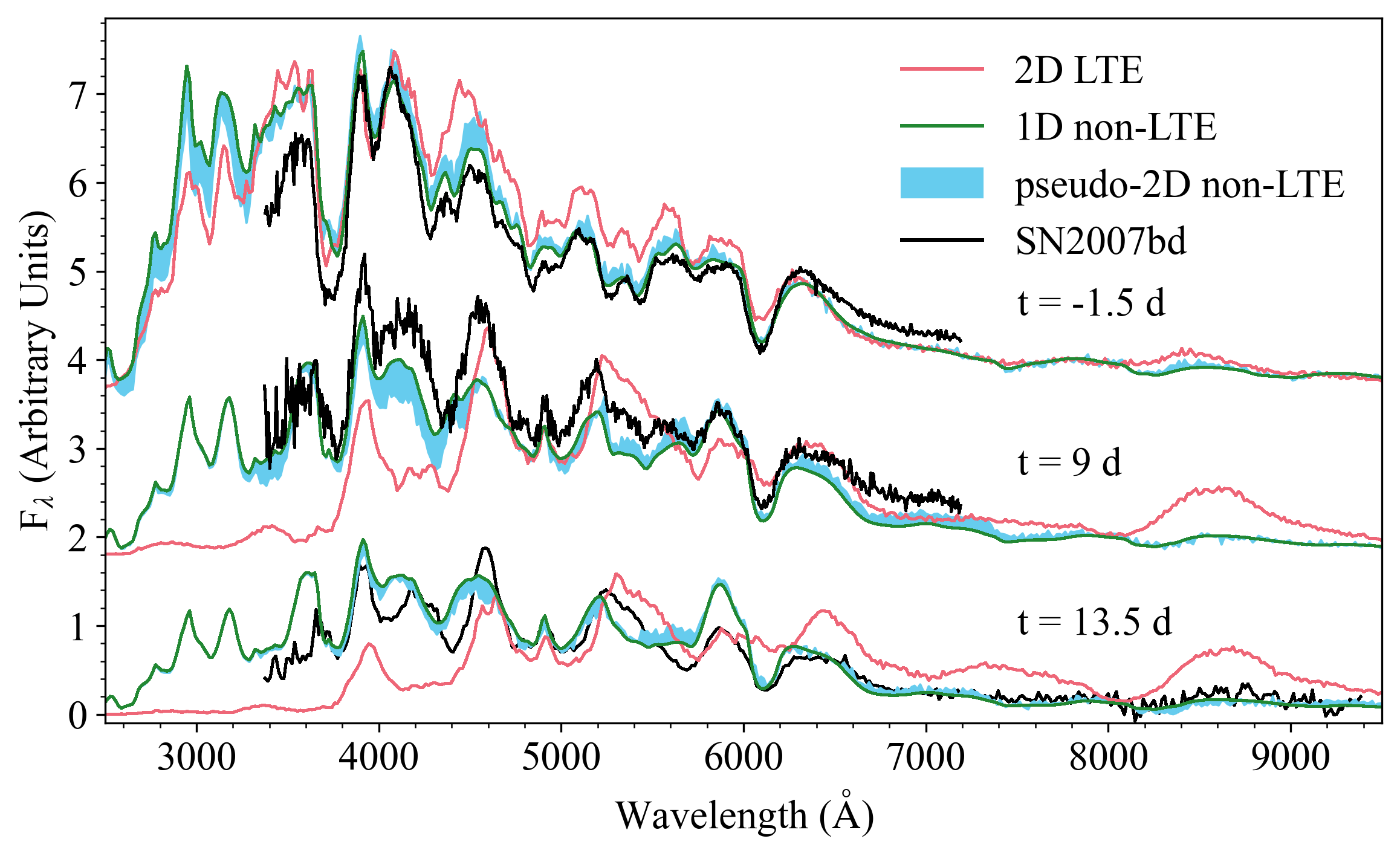}
    \caption{Spectrum of SN 2007bd \citep{blondin_2012,folatelli_2013} compared with our pseudo-2D non-LTE and 2D LTE spectra for wedge 7 of our model.
    Spectra are labeled based on the time relative to their respective $B$-band maximum.
    }
    \label{fig:2007bd_spectra}
\end{figure}

We show spectra of our pseudo-2D non-LTE and 2D LTE calculations along with SN 2007bd \citep{blondin_2012,folatelli_2013} in Figure \ref{fig:2007bd_spectra}.
Just before maximum light, the non-LTE spectrum is in modestly improved agreement with SN 2007bd, including the reproduction of Si \textsc{iii} $\lambda 4560$ at $\sim4400$ \AA.
At 9 days after maximum light, the non-LTE spectra continues to better resemble SN 2007bd.
Most of the observed spectrum is well-reproduced by the non-LTE spectrum at this time while the LTE shows poor overall agreement, especially blueward of $\sim4400$ \AA.
At 13.5 days post-maximum light, the agreement between the non-LTE and observed spectrum is less notable, but the LTE spectrum struggles even more than at earlier times, with reduced emission and increased line blanketing at low wavelengths associated with the lower temperature and ionization of the LTE calculation (see Figure \ref{fig:ionization_profiles}).

\begin{figure}
    \centering    
    \includegraphics[width=.50\textwidth]{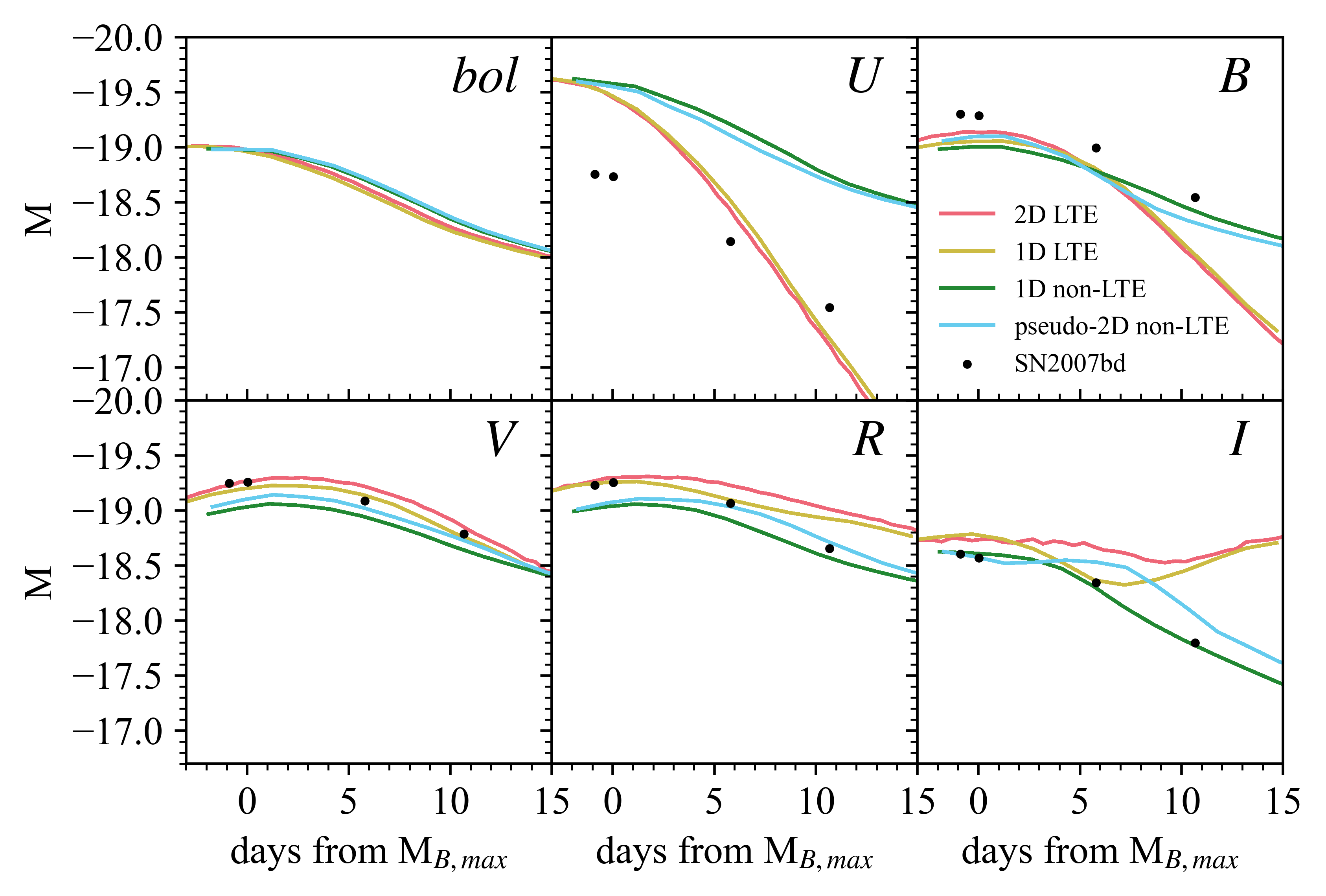}
    \caption{Light curves of SN 2007bd \citep{stritzinger_2011} compared with that from each of our radiative transfer calculations of wedge 7 of our model.
    }
    \label{fig:2007bd_lc}
\end{figure}

The light curves for SN 2007bd \citep{stritzinger_2011} are shown in Figure \ref{fig:2007bd_lc} where they are compared to each of our radiative transfer calculations of wedge 7.
The non-LTE light curves exhibit more similar shapes to SN 2007bd compared to the LTE, especially in the $I$-band where the non-LTE light curve matches the decline of the observed event 5 days after peak brightness.
Additionally, while the peak brightness in $B$-band between observation and our synthetic light curves are not in complete agreement, the non-LTE light curves replicate the shallower decline from peak.
All of our synthetic observables are much brighter in $U$-band at peak compared to SN 2007bd, however, the rate of decline in this band post-peak is much better captured by the non-LTE light curves.

\subsubsection{SN 1999ek}

\begin{figure}
    \centering    
    \includegraphics[width=0.5\textwidth]{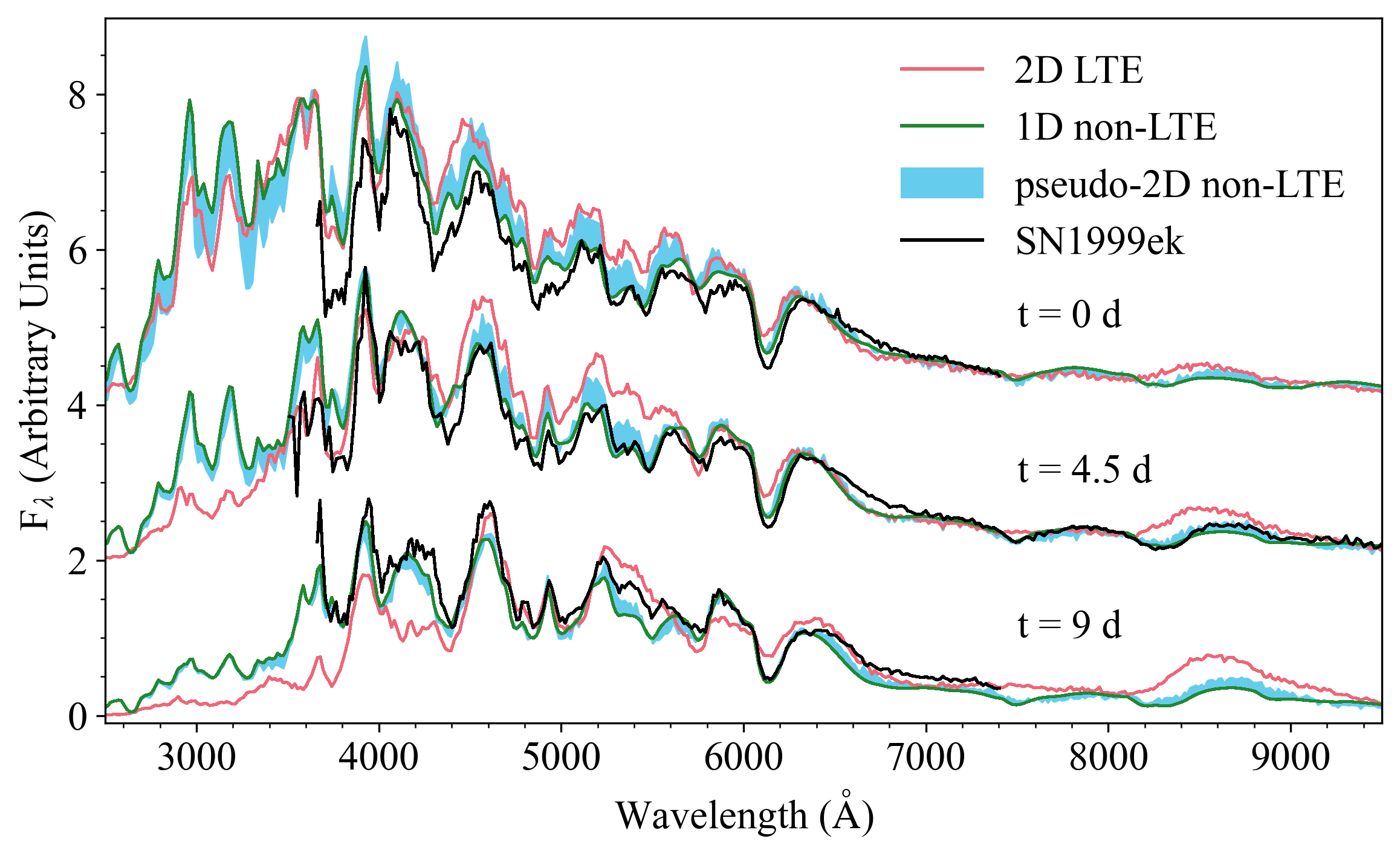}
    \caption{Spectrum of SN 1999ek \citep{blondin_2012,silverman_2012} compared with the pseudo-2D non-LTE and 2D LTE spectra of wedge 11 of our model.
    Spectra are labeled based on the time relative to their respective $B$-band maximum.
    }
    \label{fig:1999ek_spectra}
\end{figure}

We show a spectral comparison between SN 1999ek \citep{blondin_2012,silverman_2012} and both our 2D LTE and pseudo-2D non-LTE synthetic spectra for wedge 11 in Figure \ref{fig:1999ek_spectra}.
At maximum light, the LTE and non-LTE spectra shown broadly mimic the spectrum of SN 1999ek to a similar degree.
There are some differences in the two calculation's ability to reproduce the observed spectrum however, such as in the Ca \textsc{ii} H\&K line depth and shape where the non-LTE spectrum shows a local peak within the trough.
On the other hand, the non-LTE spectrum shows the Si \textsc{iii} $\lambda 4560$ feature in its maximum light spectrum that is well-matched to SN 1999ek.

At 4.5 days post-maximum light, the non-LTE and LTE observables begin to deviate in a more significant way.
It can be observed in Figure \ref{fig:1999ek_spectra} that the non-LTE spectrum always begins to replicate SN 1999ek in a stronger way overall, with a steeper spectral shape overall and deep Si \textsc{ii} $\lambda 6355$ feature.
Additionally, the absorption at 8300 \AA\ is slightly better reproduced by the pseudo-2D non-LTE spectrum.
By 9 days after maximum light, the non-LTE spectrum shows distinct regions that are more congruent with this observation of SN 1999ek at 4200 and 5400 \AA.

\begin{figure}
    \centering    
    \includegraphics[width=0.50\textwidth]{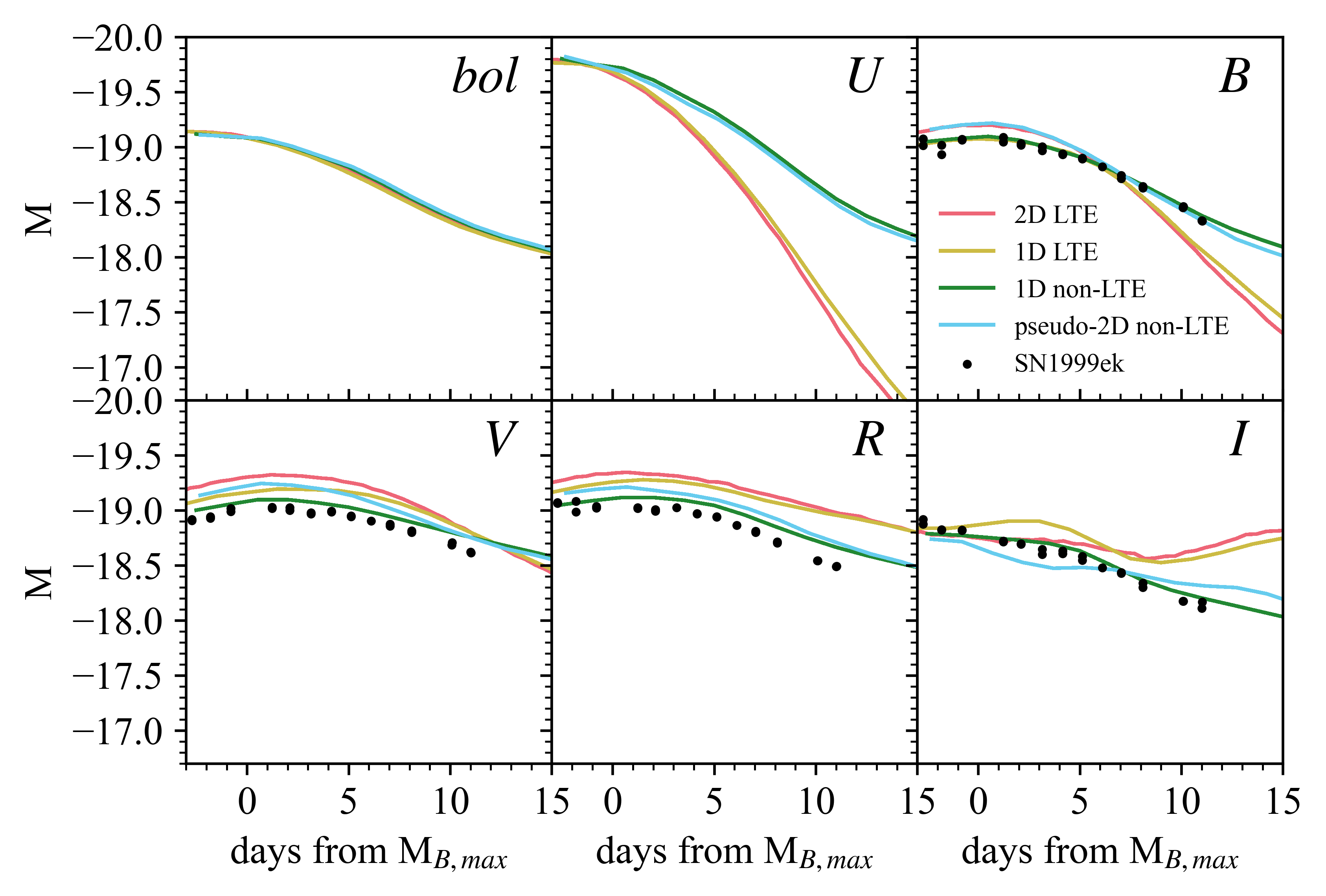}
    \caption{Bolometric and UBVRI light curves for all four sets of our synthetic observables for wedge 11, compared with photometry of SN 1999ek \citep{krisciunas_2004}.
    }
    \label{fig:1999ek_lc}
\end{figure}

In Figure \ref{fig:1999ek_lc}, we compare the light curves of wedge 11 from all of our calculations with SN 1999ek.
The peak values in the bands with photometry are more consistent with the non-LTE light curves though this agreement is sensitive to the corrections to the observed photometry, especially for host galaxy extinction.
More significantly, the non-LTE light curves show much better agreement with observation beyond peak, especially within $B$-band where decline from peak is poorly reproduced by the LTE light curves.
Correspondingly, the $I$-band light curve of SN 1999ek is adequately reproduced by the non-LTE calculations, whereas the LTE light curves see a significant departure from observation.

\subsection{Observational Correlations}
\label{subsec:obs_correlations}

\begin{figure}
    \centering    
    \includegraphics[width=0.50\textwidth]{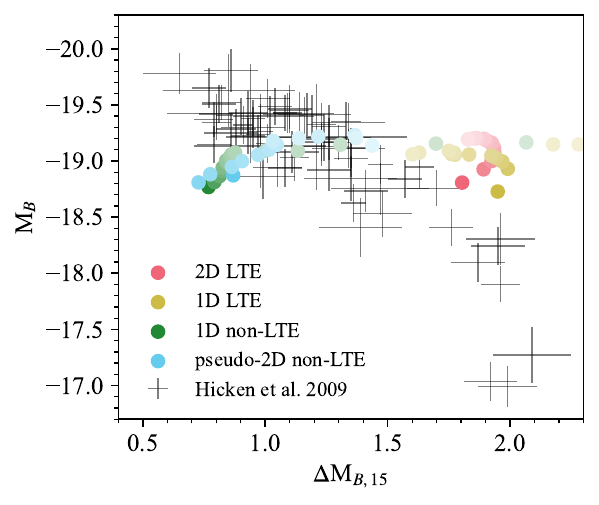}
    \caption{Phillips relation for observed SNe Ia from \citet{hicken_2009} and our four sets of calculations.
    Markers for our theoretical data have saturations based on their line of sight, with saturation decreasing from the northernmost line of sight.
    }
    \label{fig:phillips_relation}
\end{figure}

We compare the observed Phillips relation \citep{phillips_1993,hicken_2009} to that from our single 1.0 M$_{\odot}$ model across its 15 lines of sight for all four of our synthetic observable sets in Figure \ref{fig:phillips_relation}.
The observational data has been adjusted for Galactic extinction and K-corrections (see \citealt{hicken_2009}), in addition to distance and host galaxy extinction.
Both the 1D and 2D LTE calculations for this model are discrepant with the observed Phillips relation, with decline rates in the $B$-band that are appreciably larger than those seen in SNe Ia of similar peak magnitudes.
Additionally, the 1D LTE calculation shows an inability to precisely recreate the relatively smooth line of sight dependence seen in the 2D LTE.

While the extent of the peak $B$-band magnitudes remains fairly similar between the LTE and non-LTE calculations, there is a dramatic reduction in the decline rate of the $B$-band light curve at most lines of sight under non-LTE.
This shift in decline rate moves most of the lines of sight into greater agreement with observation.
Some northern lines of sight, however, show decline rates that are now slightly smaller than observed events at the same magnitude.
Additionally, the two southernmost lines of sight under non-LTE treatments have decline rates that are larger than what is expected of events of those brightnesses.
This extent of decline rates that goes beyond observed values may be a consequence of the 2D hydronuclear and radiative transfer calculations which may be exaggerating aspects of the event near the poles, motivating future 3D considerations.
We note that the multi-dimensional corrections to the non-LTE observables are somewhat modest (compare \tdnltecolor{} and \odnltecolor{} markers in Figure \ref{fig:phillips_relation}), except for the northernmost line of sight.
In summary, these non-LTE results show much greater agreement with the observed Phillips relation than that of LTE calculations, but the complete extent for this model exceeds that which is observed.

In the overall sub-Chandrasekhar scenario for SNe Ia, WD mass is the main parameter for the Phillips relation, with more massive progenitors leading to brighter and slower declining events \citep{blondin_2017,shen_2018,polin_2019,Shen_2021_nlte}.
We find for our one examined model that lines of sight with brighter $B$-band peaks decline more quickly than those that have dimmer peaks when using non-LTE.
In other words, the non-LTE Phillips relation for our individual model is nearly inverted from the observed Phillips relation.
We emphasize that the double detonation model explored in this work is of a single mass at several lines of sight.
Based on the non-trivial transformation of the observables from LTE to non-LTE (both 1D and pseudo-2D), it is challenging to precisely predict how the results from other masses, both above and below 1.0 M$_{\odot}$, will manifest on the Phillips relation.
However, it may be possible for the overall theoretical Phillips relation for a suite of masses, along with the variability imposed by line of sight diversity, to reconstruct that of the observed relation.
For example, the equatorial line of sight for a variety of masses may fit the general Phillips relation (resembling Figure 3 of \citealt{Shen_2021_nlte}), while the other lines of sight contribute to the scatter.
This line of sight-induced scatter effect has previously been demonstrated for the double detonation \citep{Shen_2021_dd,collins_2022,pollin_2024}, as well as the delayed detonation scenario for SNe Ia \citep{sim_2013}.

\begin{figure*}
    \centering    \includegraphics{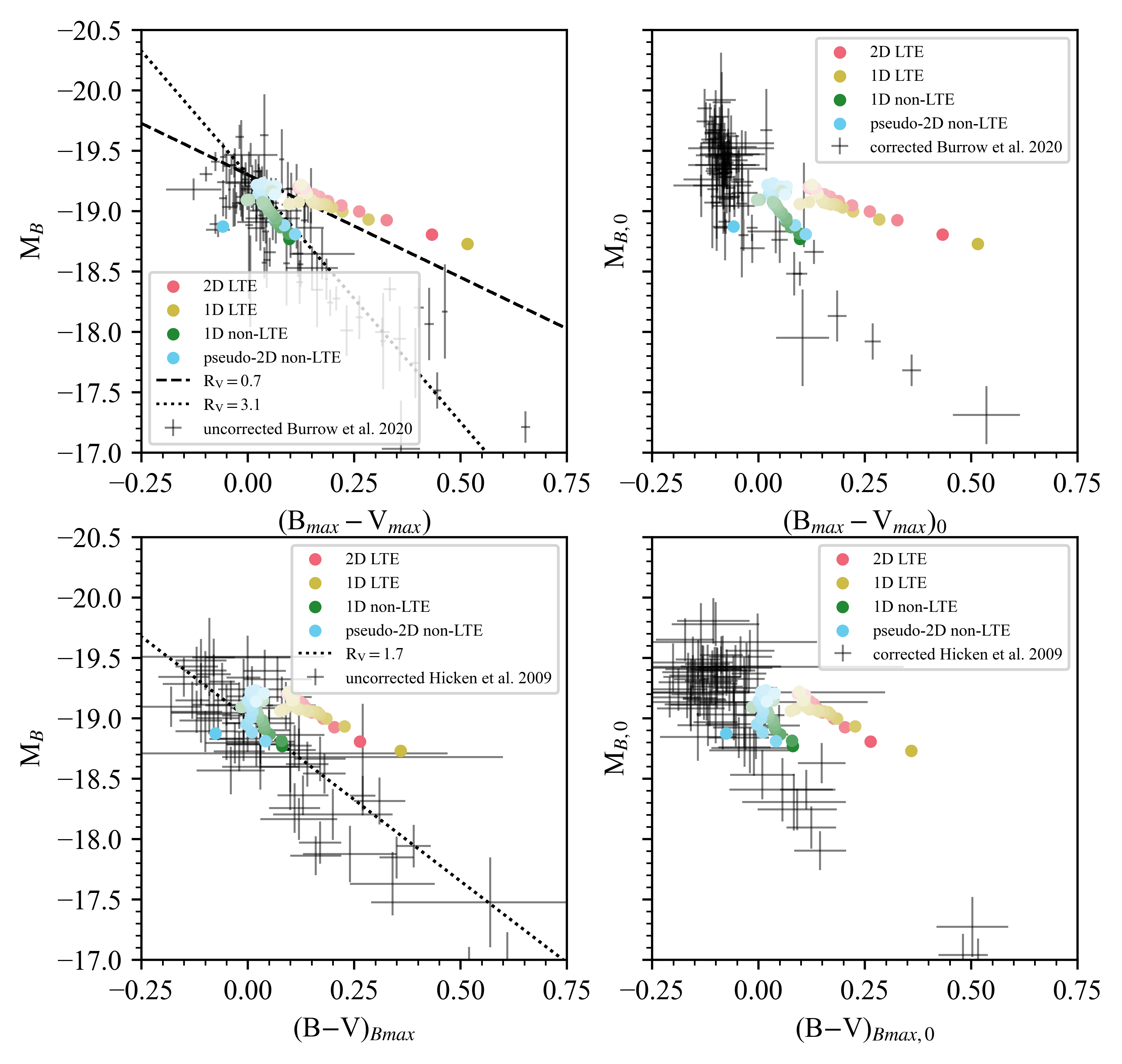}
    \caption{Color-magnitude plots for our four sets of synthetic observables, along with observational data from \citet{burrow_2020} and \citet{hicken_2009}.
    Markers for our theoretical data have saturations based on their line of sight, with saturation decreasing from the northernmost line of sight.
    The left column of plots show observed magnitudes and colors that are corrected for Galactic extinction and distance, while the right column of plots show those that are additionally corrected for host galaxy extinction.
    We also include example host galaxy correction trajectories.
    }
    \label{fig:color_magnitude}
\end{figure*}

Color-magnitude relations for observed SNe Ia and our model with the various radiative transfer
assumptions discussed above are shown in Figure \ref{fig:color_magnitude}.
We show comparisons with two varieties of color: (B$_{max}$$-$V$_{max}$) and (B$-$V)$_{Bmax}$.
All observational data shown are corrected for Galactic extinction while that shown in the right panels have also been corrected for host galaxy extinction.
Theoretical colors for our models have no dust and thus are the same on the left and right panels.
Determination of host galaxy extinction corrections is inherently model-dependent, as it must make assumptions about the intrinsic color variation of the population (or lack thereof).
Our own data on Figure \ref{fig:color_magnitude} demonstrates why this is expected to be particularly challenging, since the intrinsic color variation at different lines of sight for our non-LTE models is roughly parallel to the expected variation with different amounts of intervening dust.

Similar to previous works on the double detonation scenario \citep{polin_2019,Shen_2021_dd,collins_2022}, the color determined from LTE calculations of this double detonation model are much redder than what is observed across each of the color metrics at a given peak magnitude.
However, much greater agreement is seen between the non-LTE calculations and observations.
Our non-LTE results are congruent with the observations that have not been attempted to be corrected for host galaxy extinction and are roughly on the boundary of the population of events that have been corrected for host galaxy extinction.
Additionally, the colors of our non-LTE observables show a similar trend with peak $B$-band magnitude within the relevant magnitude range, compared to that of our LTE calculations which show a much shallower M$_B$-(B$_{max}$$-$V$_{max}$) relation.
Excluding the boundary lines of sight, we also highlight the relatively minor difference that the multidimensional effects have on the color-magnitude relation, especially for (B$_{max}$$-$V$_{max}$).

This demonstrates that, within our scenario, host galaxy extinction corrections are model dependent and so, if the double detonation model is correct, we believe that the strategy for the color correction done in the right panel in Figure \ref{fig:color_magnitude} will need changes.
It seems likely that some portion of the color differences among observations that have been attributed to differences in dust column may arise from differences in intrinsic color instead.
A much wider set of models will be needed before firm inferences can be drawn.

\begin{figure}
    \centering    
    \includegraphics[width=0.50\textwidth]{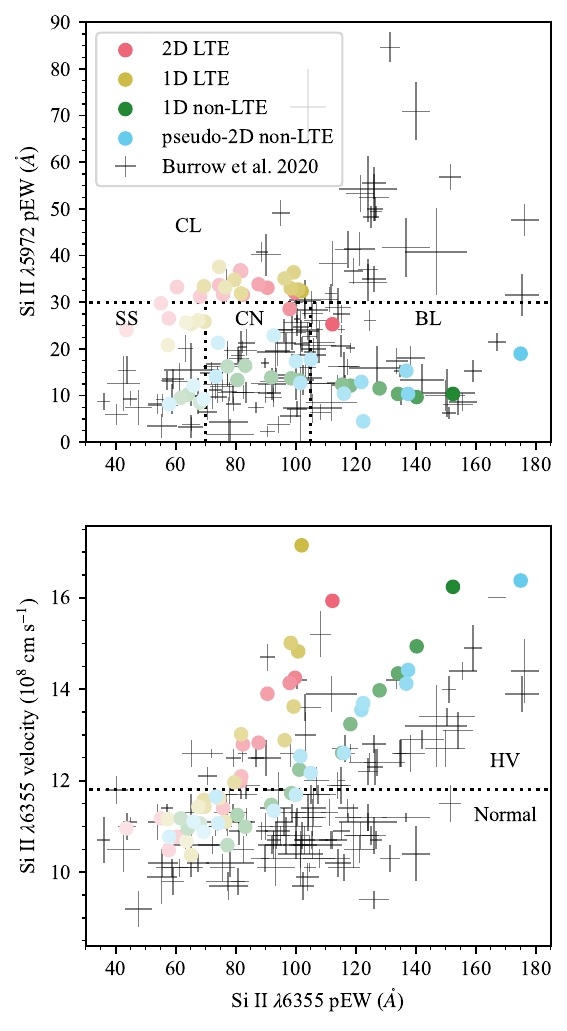}
    \caption{
    Spectral feature comparison for our models and observed SNe Ia from \citet{burrow_2020}.
    Both observational and theoretical values shown are calculated using \texttt{Spextractor} (\citet{papadogiannakis_2019}, modified by \citet{burrow_2020}).
    The Branch classes \citep{branch_2006} are labeled and delineated by a dotted line in the top panel, based on approximate boundary values from \citet{folatelli_2013}.
    In the bottom panel, the Wang classes \citep{wang_2009} are labeled and separated by a dotted line.
    Spextractor did not detect a Si \textsc{ii} $\lambda$5972 pEW in the pseudo-2D maximum light spectra of wedge 1, so it is integrated manually.
    }
    \label{fig:sii_vel}
\end{figure}

We show in the upper panel of Figure \ref{fig:sii_vel} how a few spectral indicators from our set of model observables fare in the Branch \citep{branch_2006} classification of SNe Ia.
Observed events are classified in the Branch system based on the pEW of two ubiquitous absorption features at maximum light, Si \textsc{ii} $\lambda$5972 and Si \textsc{ii} $\lambda$6355.
A majority of observed SNe Ia fall within a small Si \textsc{ii} $\lambda$5972 pEW region, where they are sorted into shallow-silicon (SS), core-normal (CN), and broad-line (BL) classes based on Si \textsc{ii} $\lambda$6355 pEW in an increasing order.
We note that these classes are relatively continuous and are not necessarily considered to be distinct groups of events.
The 1D and 2D LTE calculations of our one model straddle the boundary between these groups and the cool (CL) class of SNe Ia, specifically at a region in this parameter space that has a dearth of observed events.
In the non-LTE treatments of the synthetic observables, however, the Si \textsc{ii} $\lambda$5972 pEWs are consistently lower at all lines of sight and are brought into much better agreement with a portion of observed SNe Ia for these spectral indicators.
The non-LTE spectra span three Branch classes, with the southern, equatorial, and northern lines of sight representing the SS, CN, and BL classes, respectively.
This may be an indication that these identified classes of SNe Ia can be at least partially attributed to line of sight.
We also find that the extent of the Si \textsc{ii} $\lambda$6355 pEW is wider in our non-LTE treatment of the synthetic observables compared to LTE, nearly spanning the breadth from that of observed SNe Ia.

In the lower panel of Figure \ref{fig:sii_vel}, we show the pEW and velocity of Si \textsc{ii} $\lambda$6355 for our models and observed SNe Ia.
The calculated Si \textsc{ii} $\lambda$6355 velocity is fairly consistent between the non-LTE and LTE treatments for each line of sight, with more northern lines of sight showing higher velocities.
We delineate this plot based on the Wang classification \citep{wang_2009} that categorizes observed events with Si \textsc{ii} $\lambda$6355 velocity above $\sim11,000$ km s$^{-1}$ as high velocity (HV).
\citet{wang_2009} shows that HV events are redder on average, either intrinsically or otherwise.
Interestingly, our model spectra that could be classified as HV occur at northern lines of sight, which also see redder colors (see Figure \ref{fig:color_magnitude}), another indication that some spectral classes may be able to be explained with line of sight effects.
The normal and HV Wang classes on average do not have distinctly different peak magnitudes as our individual model does across lines of sight, however.
The characteristics shown in Figure \ref{fig:sii_vel} are modestly more impacted by dimensionality than what was shown previously for the Phillips relation and color-magnitude, likely due to the increased sensitivity of these parameters.


\section{Conclusions}
\label{sec:conclusions}

Here we have described a technique to generate non-LTE observables from a multidimensional SN Ia model, motivated by the fact that spherically symmetric models and LTE calculations are insufficient to produce observables that match observation to a precise degree.
We show that 1D steady-state calculations reproduce, to good approximation, the results from a time-dependent 2D calculation along corresponding lines of sight.
As a result, this makes it possible to produce reasonably approximate multidimensional non-LTE spectra using 1D non-LTE calculations along individual lines of sight.
An important component of this strategy is the constraint of the outgoing bolometric luminosity in each of the 1D calculations to match that of the 2D LTE calculation at corresponding lines of sight, which is necessary primarily due to the varying amounts of radioactive material in the ejecta profiles for different lines of sight.
This technique is important as non-LTE calculations are essential for the calculation of a number of SNe Ia features and relations, including the $B$-band Phillips relation, but are often prohibitively expensive if not reduced in accuracy in multiple dimensions.
In the double detonation model, for example, the asymmetry of the ejecta is an essential aspect as it allows for a variety of observable indicators based on the viewing angle to a given explosion, including peak magnitude and expansion velocity, which also vary among the population of observed SNe Ia.
This technique allows for high-accuracy non-LTE observables from relatively efficient calculations while preserving the asymmetry that is inherent to the model, improving the unification between theory and observation.

This work demonstrates the importance of using non-LTE physics to determine synthetic observables from SNe Ia models, especially beyond maximum light, as we have shown several notable changes to the observables due to the non-LTE treatment which are in much greater excess than the effects from multidimensionality.
Overall, the non-LTE synthetic observables presented in this work show much greater agreement with observation for this double detonation model compared to the LTE observables especially at times much beyond maximum light.
This includes bluer colors, the production of Si \textsc{iii} $\lambda 4560$ around maximum light, significant improvement of the spectra many days post-peak, shallower $B$-band declines, and a reduction of the Si \textsc{ii} $\lambda$5972 pEW.
These improvements benefit comparisons between individual lines of sight and observed events, as well as how the examined model across all lines of sight compares to correlations of SNe Ia populations.

While the observables highlighted in this work show general improvement, issues still remain for this model, particularly at northern lines of sight.
This includes Si \textsc{ii} $\lambda$6355
velocities and B-V color that are somewhat higher than observed, in addition to slightly small $\Delta M_{B,15}$ values.
It is possible that these issues may be alleviated by 3D modeling of the double detonation scenario as the polar wedges from our 2D models are most affected by the imposed cylindrical symmetry.
Additionally, effects from the companion star, which are not considered in the model examined in this paper, can alter the amount of diversity across lines of sight \citep{boos_2024,pollin_2024}.

Future work on this technique will involve a closer examination of the origin of the multidimensional deviations of the observables (e.g.\ Figure \ref{fig:1d_vs_2d_lte_spectra}) to determine the quality of approximation of this strategy.
Additionally, it would be fruitful to establish avenues to reduce the uncertainty brought upon by using different codes between the non-LTE and LTE calculations.
The disparity between the LTE (Sedona) and non-LTE (\texttt{CMFGEN}) results in this work are qualitatively similar to what is shown by \citet{blondin_2022}, which reviewed the consistency between several radiative transfer codes for SNe Ia.
As different radiative transfer codes can produce varied results, using a standardized method that can isolate specific physical processes, while still considering multidimensional effects, would improve the reliability and interpretation of the results generated from our technique.

We will also use this technique on a wider range of models.
This includes double detonation models with masses above and below 1.0 M$_{\odot}$, in addition to double detonation models that involve a companion.
Non-LTE observables from models that include an exploding companion (i.e.\ triple or quadruple detonation;  \citealt{papish_2015,tanikawa_2019,Pakmor_2022,pollin_2024}) will also be particularly interesting due to recent work that has shown, in LTE, that events arising from one or two star explosions in this scenario can have remarkably similar spectra \citep{Pakmor_2022,boos_2024,pollin_2024}.
Other sub-Chandrasekhar SN Ia models may also be used with this technique to determine how they fare in comparison to the population of observed SNe Ia.

\begin{acknowledgments}

We thank the referee for their helpful comments.
We also thank Stefan Taubenberger for his reduction of the \citet{hicken_2009} data.

S.J.B.\ and D.M.T.\ acknowledge support from the National Science Foundation under Grant No.\ 2307442.
S.J.B.\ also acknowledges support from NASA grant HST AR-16156.
K.J.S.\ is supported by NASA through the Astrophysics Theory Program (80NSSC20K0544).

This work was granted access to the HPC resources of TGCC under the allocation 2023 -- A0150410554 on Irene-Rome made by GENCI, France. 
Resources supporting this work were also provided by the NASA High-End Computing (HEC) Program through the NASA Advanced Supercomputing (NAS) Division at Ames Research Center and the University of Alabama high performance computing facility.

\software{
\texttt{Sedona} (\citealt{kasen_2006}),
\texttt{CMFGEN} (\citealt{hillier_2012}),
\texttt{Spextractor} (\url{github.com/anthonyburrow/spextractor}),
\texttt{SNooPy} (\citealt{Burns_2011}),
\texttt{matplotlib} (\citealt{matplotlib}, \url{matplotlib.org}).
}

\end{acknowledgments}

\appendix
\section{Full Spectral Sets}

\begin{figure*}
    \centering    \includegraphics[width=1.0\textwidth]{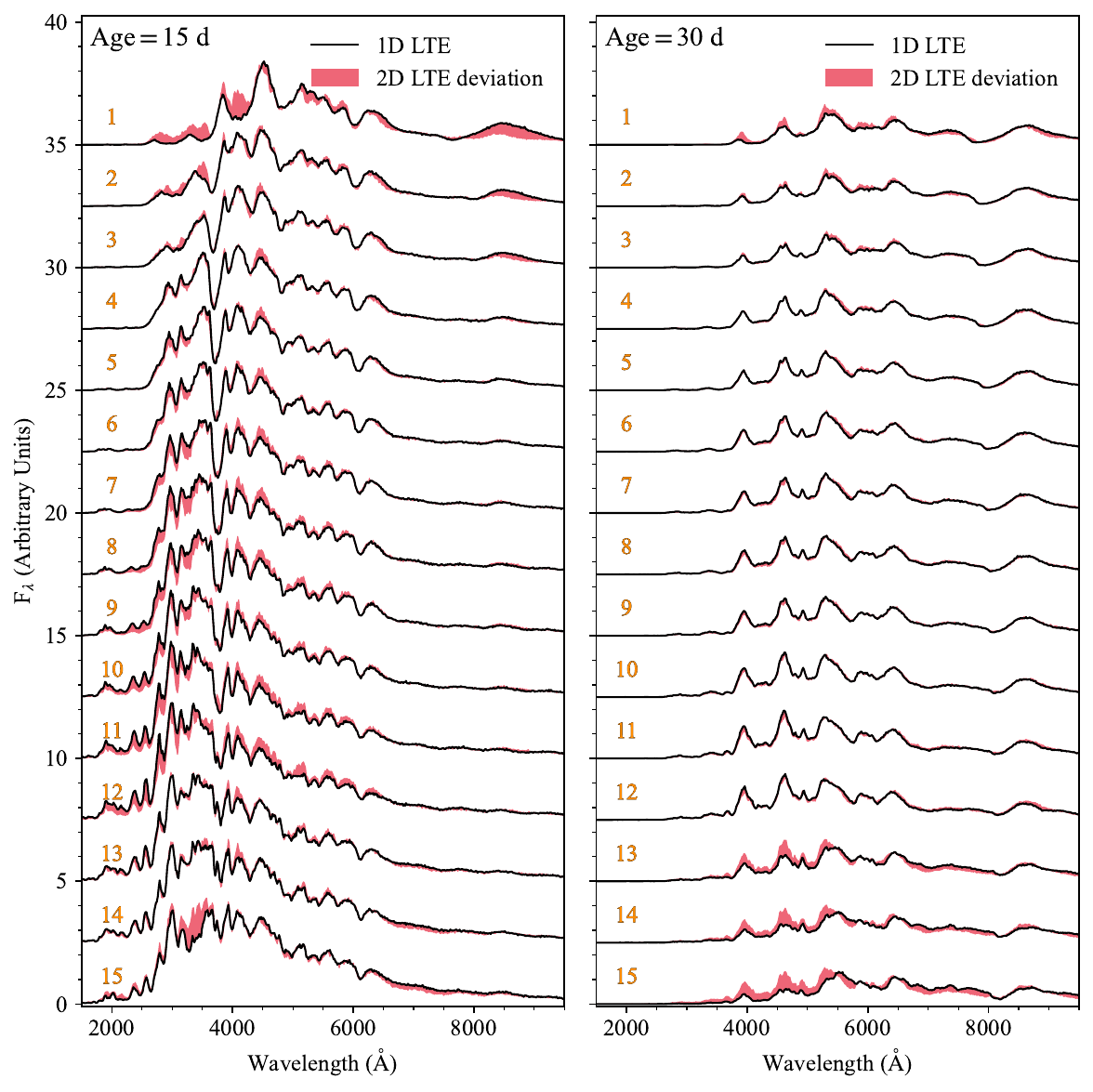}
    \caption{LTE spectra at all lines of sight at 15 and 30 days after the explosion.
    Spectra from our 1D LTE calculations are shown in black while the \tdltecolor{} shading indicates how the 2D LTE spectra deviate from the 1D LTE spectra (i.e.\ the outer edge of the \tdltecolor{} shading is the 2D spectra).
    The lines of sight are labeled in orange, corresponding with the wedges in Figure \ref{fig:boos21_ejecta}.
    }
    \label{fig:appendix_lte}
\end{figure*}

\begin{figure*}
    \centering    
    \includegraphics[width=1.0\textwidth]{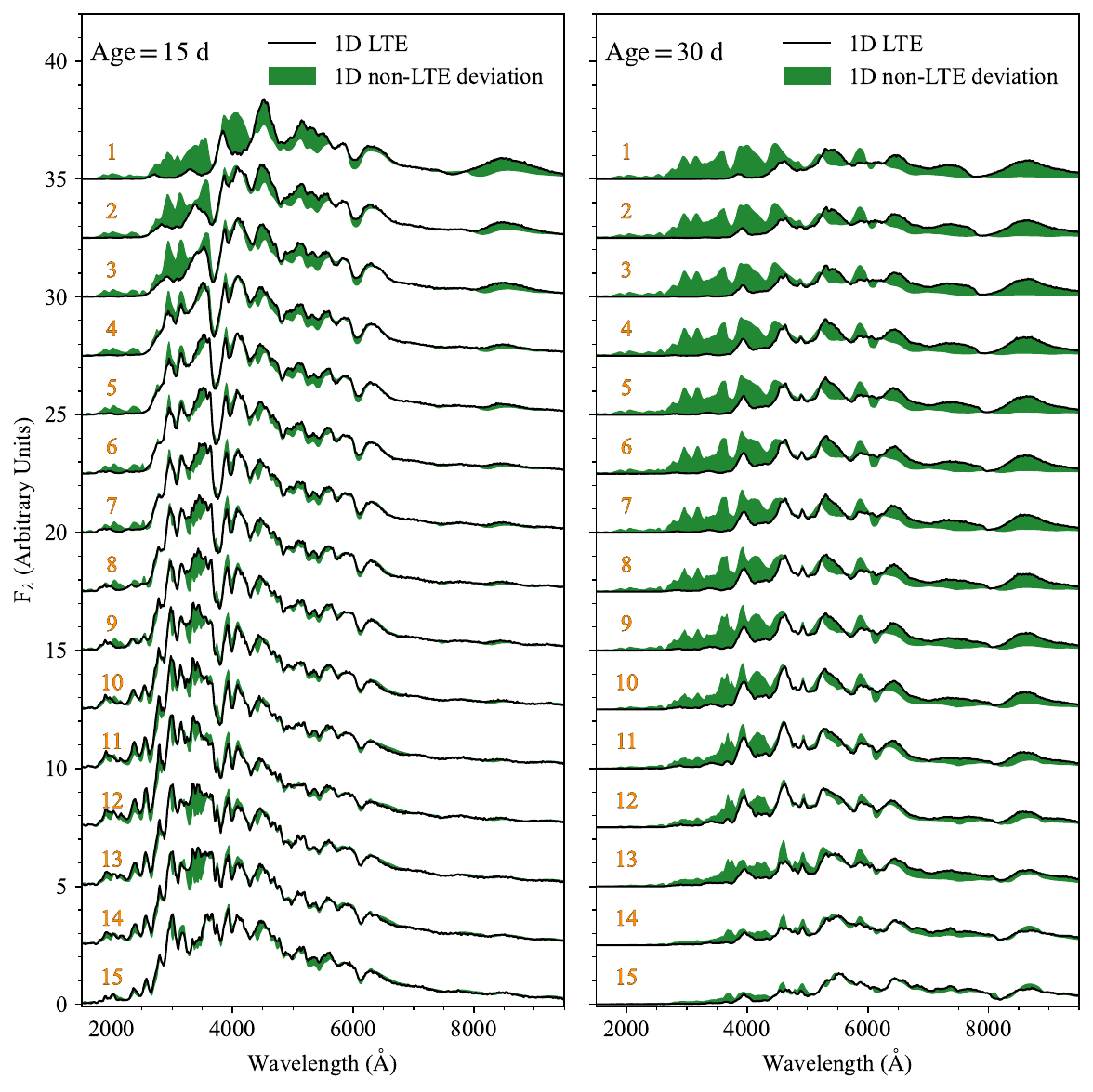}
    \caption{1D spectra at all lines of sight at 15 and 30 days after the explosion.
    Spectra from our 1D LTE calculations are shown in black while the \odnltecolor{} shading indicates how the 1D non-LTE spectra deviate from the 1D LTE spectra (i.e.\ the outer edge of the \odnltecolor{} shading is the 1D non-LTE spectra).
    The lines of sight are labeled in orange, corresponding with the wedges in Figure \ref{fig:boos21_ejecta}.
    }
    \label{fig:appendix_nlte}
\end{figure*}

We show the 1D LTE spectra for all wedges at 15 and 30 days post-explosion, compared with the 2D LTE and 1D non-LTE spectra in Figures \ref{fig:appendix_lte} and \ref{fig:appendix_nlte}, respectively.

\section{NIR Spectra}
Here we highlight the IR spectra from our calculations, in comparison with SN 2011fe.
The observed spectrum shown is corrected for distance, redshift, and Galactic reddening.
The 2D LTE and pseudo-2D non-LTE synthetic spectra are rebinned beyond 10,000 \AA\ due to the poor Monte-Carlo sampling at these wavelengths.

\begin{figure*}
    \centering    
    \includegraphics[width=0.5\textwidth]{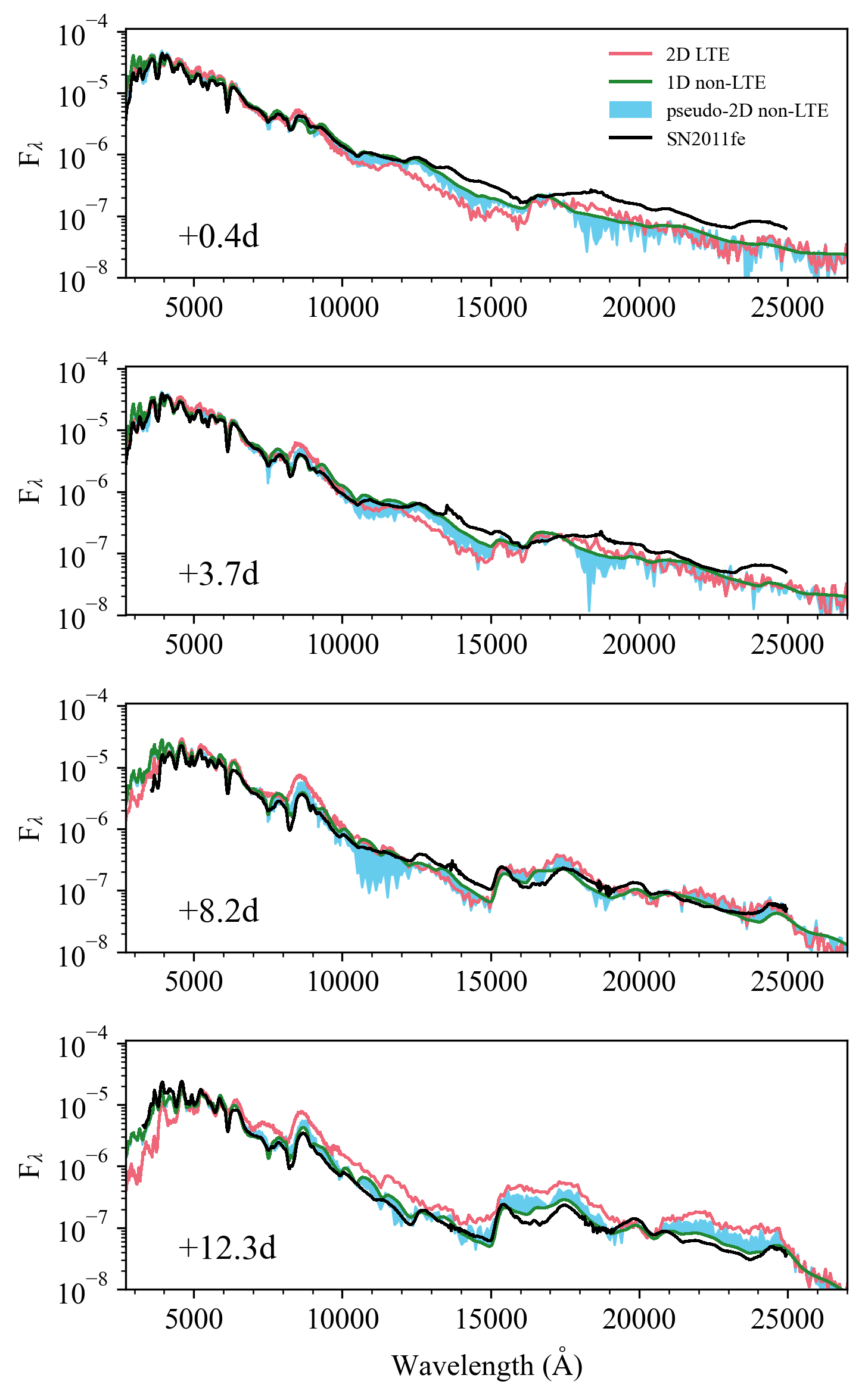}
    \caption{Optical-NIR spectrum of SN 2011fe \citep{Hsiao_2013} compared with our 1D non-LTE and 2D LTE spectra for wedge 12 of our model.
    The \tdnltecolor{} shading indicates the difference between 1D non-LTE and pseudo-2D non-LTE spectra at the given time.
    Spectra are labeled based on the time relative to their respective $B$-band maximum.
    }
    \label{fig:appendix_IR_2011fe}
\end{figure*}

We show an NIR spectral comparison between SN 2011fe and that from wedge 12 of our calculations in Figure \ref{fig:appendix_IR_2011fe}.
Around maximum light, the 1D non-LTE spectra provides a satisfactory match with SN 2011fe in the NIR out to $\sim13000$ \AA, which includes a number of Ca \textsc{ii} and Mg \textsc{ii} features (see \citealt{Hsiao_2013} for a detailed discussion of the NIR spectral features of SN 2011fe).
However, the modest pseudo-2D non-LTE deviation from the 1D non-LTE suggests that multidimensional effects may be important in the NIR.
Beyond $13000$ \AA, none of our model spectra are able to well-replicate the flux or features observed in SN 2011fe at max light, or 3.7 days post-max.
By 8.2 days, however, generally improved agreement can be observed in most of the NIR region displayed between SN 2011fe and the 1D non-LTE spectra.
The improved agreement between SN 2011fe and the 1D non-LTE calculation is more stark at 12.3 days, but the pseudo-2D non-LTE is notably deviated from the 2D LTE beyond $15000$ \AA.

\section{Alternative Power Injection Technique}
\label{sec:alt_power}
An alternative method to increase the outgoing luminosity of the 1D LTE calculations to match that of the 2D LTE treatment was used in a previous version of this work, which we describe here. 
In this method, rather than scaling the radioactive decay energies, we introduce an artificial power source at the origin of the grid.
The luminosity of this source is set such that the total outgoing luminosity of a given calculation in 1D matches the 2D results of the corresponding line of sight.
The source is a blackbody, with the results being sensitive to the choice of temperature ($3\times10^4$ K is used here).
Additional testing showed that the power source could be placed out to upwards of $\sim5,000$ km s$^{-1}$ without significant alterations to the spectra.

\begin{figure*}
    \centering    
    \includegraphics[width=1.0\textwidth]{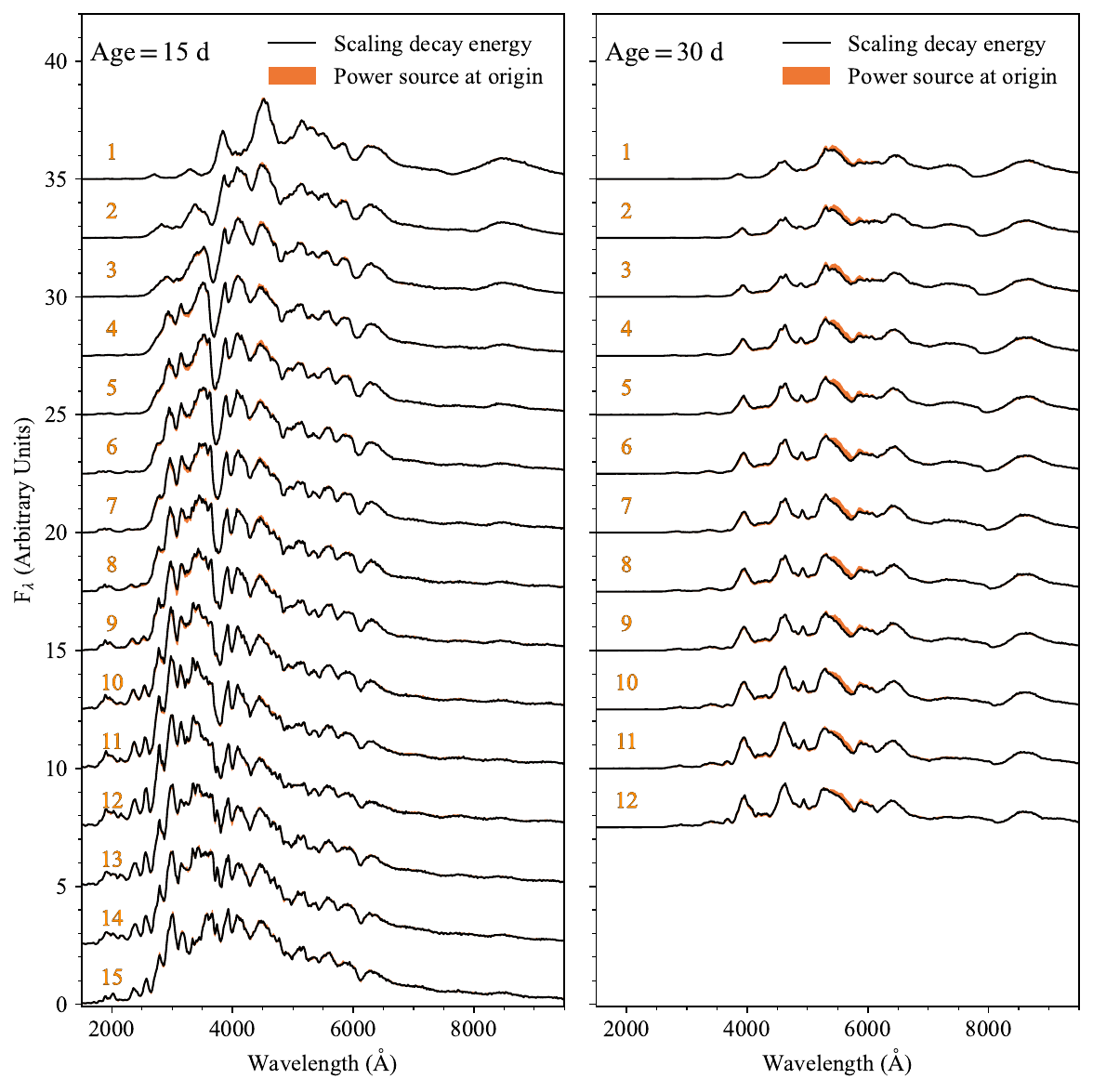}
    \caption{1D spectra at all valid lines of sight at 15 and 30 days after the explosion.
    Spectra from our standard set of 1D LTE calculations, where the decay energy is scaled, are shown in black.
    The orange shading indicates how spectra under the alternative method, where a power source is placed at the origin, deviates from our standard strategy.
    The lines of sight are labeled in orange, corresponding with the wedges in Figure \ref{fig:boos21_ejecta}.
    The three southernmost lines of sight are not shown at 30 days post-explosion as their luminosities are unable to be unified.}
     \label{fig:appendix_spectra_alt_power}
\end{figure*}

The spectra of the two 1D LTE methods are compared in Figure \ref{fig:appendix_spectra_alt_power}.
At 15 days post-explosion, the two strategies produce nearly identical spectra.
By 30 days post-explosion, the spectra are still broadly very similar for most wedges, except for the region around 5500 \AA.
Our decay energy scaling method is more effective in reproducing the 2D spectra in this region at this time (see Figure \ref{fig:appendix_lte}).
Spectra at 30 days post-explosion for wedges 13-15 are not shown in Figure \ref{fig:appendix_spectra_alt_power} as these spectra have an unmodified bolometric luminosity in steady-state that is higher than the target 2D result, with no functional way to reduce the outgoing luminosity in the power source strategy.

The difference in physical motivation and effectiveness between these two techniques is fairly marginal.
However, the decay energy scaling strategy has an advantage over the other because of its ability to reduce the outgoing luminosity in a straightforward way (i.e.\ a decay energy scaling factor below unity).
This is necessary for the unification of bolometric luminosities with the 2D calculation for the three southernmost wedges at later times, which cannot be obtained through the power source method as it can only increase the outgoing luminosity.

\section{Data}
We provide the 2D ejecta and 15 wedges profiles associated with the our examined model on Zenodo (\href{https://doi.org/10.5281/zenodo.15122976}{10.5281/zenodo.15122976}).
Also included in this repository is our synthetic observables from each of our calculations (2D LTE, 1D LTE, and 1D non-LTE), as well as our pseudo-2D non-LTE observables.
This includes spectra and UBVRI light curves between -3 and +15 d from $B$-band peak, with a cadence of 1.5 d.


\bibliography{ref}{}
\bibliographystyle{aasjournalv7}

\end{document}